\documentclass[journal]{IEEEtran}
\usepackage{amsmath}
\usepackage{amssymb}
\usepackage{amsfonts}
\usepackage{algorithm}
\usepackage{algpseudocode}
\usepackage{dsfont}
\usepackage{float}
\usepackage{cite}
\usepackage{graphicx}
\usepackage{epsfig}
\usepackage{subfigure}
\usepackage{psfrag}
\usepackage{xcolor}
\usepackage{url}
\usepackage[colorlinks,linkcolor=black,urlcolor=black,anchorcolor=black,citecolor=black,hyperfootnotes=true]{hyperref}
\usepackage{bm}

\newtheorem{theorem}{Theorem}

\newtheorem{remark}{Remark}

\newcommand\blfootnote[1]{%
  \begingroup
  \renewcommand\thefootnote{}\footnote{#1}%
  \addtocounter{footnote}{-1}%
  \endgroup
}

\begin{document}
\title{Reducing Channel Estimation and Feedback Overhead in IRS-Aided Downlink System: A Quantize-then-Estimate Approach}
\author{Rui Wang, Zhaorui Wang, Liang Liu, Shuowen Zhang, and Shi Jin
\thanks{Manuscript received March 16, 2024; revised July 15, 2024 and October 19, 2024; accepted November 10, 2024. This work was supported in part by the National Key Research and Development Project of China under Grant 2022YFB2902800; in part by the National Natural Science Foundation of China under Grant 62101474; in part by the Research Grants Council, Hong Kong, China, under Grant 15203222 and 15230022; in part by the Basic Research Project under Grant HZQB-KCZYZ-2021067 of Hetao Shenzhen-Hong Kong Science and Technology Cooperation Zone. \textit{(corresponding author: Liang Liu)}}
\thanks{R. Wang, L. Liu, and S. Zhang are with the Department of Electrical and Electronic Engineering, The Hong Kong Polytechnic University, Hong Kong SAR, China (e-mails: rui-eie.wang@connect.polyu.hk, \{liang-eie.liu, shuowen.zhang\}@polyu.edu.hk).}
\thanks{Z. Wang is with the FNii and SSE, The Chinese University of Hong Kong, Shenzhen, Shenzhen 518172, China (e-mail: wangwang2020channel@cuhk.edu.cn).}
\thanks{S. Jin is with the National Mobile Communications Research Laboratory, Southeast University, Nanjing 210096, China (e-mail: jinshi@seu.edu.cn).}
}
\maketitle

\begin{abstract}
    Channel state information (CSI) acquisition is essential for the base station (BS) to fully reap the beamforming gain in intelligent reflecting surface (IRS)-aided downlink communication systems. Recently, \cite{wang2020channel} revealed a strong correlation in different users' cascaded channels stemming from their common BS-IRS channel component, and leveraged such a correlation to significantly reduce the pilot transmission overhead in IRS-aided uplink communication. In this paper, we aim to exploit the above channel property to reduce the overhead for both pilot and feedback transmission in IRS-aided downlink communication. Note that in the downlink, the distributed users merely receive the pilot signals containing their own CSI and cannot leverage the correlation in different users' channels, which is in sharp contrast to the uplink counterpart considered in \cite{wang2020channel}. To tackle this challenge, this paper proposes a novel ``quantize-then-estimate'' protocol in frequency division duplex (FDD) IRS-aided downlink communication. Specifically, the users quantize and feed back their received pilot signals, instead of the estimated channels, to the BS. After de-quantizing the pilot signals received by all the users, the BS estimates all the cascaded channels by leveraging their correlation, similar to the uplink scenario. Under this protocol, we manage to propose efficient user-side quantization and BS-side channel estimation methods. Moreover, we analytically quantify the pilot and feedback transmission overhead to reveal the significant performance gain of our proposed scheme over the conventional ``estimate-then-quantize'' scheme.
\end{abstract}

\blfootnote{The materials in this paper have been presented in part at the IEEE Global Communications Conference, December 2023 \cite{globecom2023}.}
\section{Introduction}

\subsection{Motivation}

Intelligent reflecting surface (IRS) has been recognized as a promising technique to enhance the capacity and coverage of the future 6G cellular networks, thanks to its ability to tune the channels to be favorable for communication. To design the best propagation conditions via the IRS, channel state information (CSI) acquisition is of paramount importance. However, such a task is challenging due to the vast number of channel coefficients associated with the IRS \cite{Basar2019wireless,wu2020towards,wu2021intelligent,zheng2022survey}.

This paper considers IRS-assisted downlink communication in a frequency division duplex (FDD) system, where a multi-antenna base station (BS) needs to know the BS-IRS-user cascaded channels of all the users for designing its own and the IRS's beamforming vectors. In time division duplex (TDD) systems, the downlink CSI can be acquired by estimating the uplink CSI thanks to channel reciprocity. In our considered FDD systems, the channel reciprocity does not hold and the users have to feed back some useful information from their received pilot signals to let the BS acquire the CSI. Under the conventional systems without the IRS, the ``estimate-then-quantize'' scheme \cite{jindal2006mimo,love2008overview,shen2018channel,Wen2018deep} was widely used for downlink channel estimation and feedback, where each user first estimates its downlink channels with the BS based on its received pilot signals and then feeds back the estimated channels to the BS. However, the overall overhead of this classic protocol is unaffordable in IRS-aided systems due to the following reasons. First, to enable each user to estimate a huge number of coefficients in its BS-IRS-user channel, the BS has to send a long pilot sequence, leading to high channel estimation overhead. Second, after the channel estimation phase, each user has to feed back a huge number of quantization bits for transmitting the estimated channel coefficients to the BS, leading to high feedback overhead. This calls for some innovative protocol to replace the conventional ``estimate-then-quantize'' protocol in IRS-assisted downlink systems for low-overhead communication.

\subsection{Prior Work}

Previous research has been conducted to reduce channel estimation overhead in IRS-assisted uplink communication systems, by utilizing multi-user channel correlation \cite{wang2020channel,chen2023channel}, two-timescale property \cite{hu2021two}, beamspace channel sparsity \cite{Wei2021partII,Peng2023twostage,Nadeem2020intelligent,Wei2021channel,zhou2022channel,chen2023channel}, IRS elements grouping \cite{zheng2020intelligent}, etc. In TDD downlink systems, the above methods can be used for the BS to obtain the CSI based on channel reciprocity. However, in our considered FDD downlink systems, dedicated methods should be proposed for low-overhead channel estimation and feedback.

Under FDD IRS-assisted downlink communication systems, most prior works are under the ``estimate-then-quantize'' scheme. In particular, they are interested in reducing the overhead in the feedback phase, assuming the channels are already estimated by the users \cite{chen2021adaptive,Ge2023beamforming,chen2023customize,Guo2023deep,shen2021dimension,Shi2022Triple}. For the single-user system, \cite{chen2021adaptive} proposed a novel cascaded codebook for the BS-IRS and IRS-user subchannel, respectively, that is synthesized by two sub-codebooks whose codewords are cascaded by line-of-sight and non-line-of-sight components. It is demonstrated that the cascaded codebook outperforms the naive random vector quantization codebook with lower feedback bits
requirement.\cite{Ge2023beamforming} reduced feedback overhead by selecting several dominant BS-IRS-user cascaded channel paths based on their contributions to spectral efficiency, instead of feeding back CSI of all the cascaded paths. However, these two works only focused on channel path gain information feedback, without considering path angle information. \cite{chen2023customize} proposed to customize a cascaded channel with a reduced number of paths for the sub-6 GHz rich scattering environment by path selection and multi-IRS phase shifter design, thus reducing the number of feedback parameters.
\cite{Guo2023deep} leveraged the two-timescale property of the cascaded BS-IRS-user channels \cite{hu2021two} to build a neural network consisting of large and small timescale feedback. The BS-IRS channel, which is assumed to be unchanged for a long time, only needs to be fed back in each large timescale, and the IRS-user channel, which changes frequently but with low dimension, is fed back in each small timescale. For the multi-user system, \cite{shen2021dimension} exploited the single-structured sparsity of the cascaded BS-IRS-user channel, i.e., in the hybrid spatial-angular domain channel matrix, all users share the same indices of non-zero columns, due to the common sparse BS-IRS channel. These user-independent non-zero columns' indices are fed back by only one user, and the user-specific non-zero column vectors are fed back by different users, thus reducing feedback overhead. In addition to the common indices of non-zero columns in the beamspace channel, \cite{Shi2022Triple} found that the non-zero values in different non-zero columns have the same location offsets and amplitude ratios. These offsets and amplitudes
are shared among all users and can be exploited to further reduce the number of feedback parameters.

Downlink CSI acquisition consists of both the downlink pilot transmission phase and the uplink feedback transmission phase. On one hand, the above works all assumed that each user knows its cascaded channels perfectly. However, due to the huge number of IRS reflecting elements, the overhead for the users to estimate their channels is huge, which is not considered in the above works. On the other hand, from feedback overhead perspective, there is a potential to significantly improve the performance of the schemes proposed in \cite{chen2021adaptive,Ge2023beamforming,chen2023customize,Guo2023deep,shen2021dimension,Shi2022Triple}. Recently, \cite{wang2020channel} revealed a unique property of the BS-IRS-user cascaded channels among different users — due to the common BS-IRS channel to all the users, each user's cascaded channel vector is a scaled version of another user's cascaded channel vector. However,  \cite{chen2021adaptive,Ge2023beamforming,chen2023customize,Guo2023deep,shen2021dimension,Shi2022Triple} did not consider how to exploit this property to reduce the feedback overhead.

\subsection{Main Contributions}

This paper aims to significantly reduce the overhead of both the pilot transmission phase and the feedback transmission phase for CSI acquisition in IRS-assisted downlink communication. Actually, the CSI acquisition problem is essentially a distributed source coding (DSC) problem \cite{DSC}, where the core question is what information should be fed back by the users after they receive the pilot signals from the BS, such that the BS can acquire the CSI with the shortest pilot signals transmitted in the downlink and the minimum quantization bits transmitted in the uplink. Due to the correlation in cascaded channels among different users revealed in \cite{wang2020channel}, the conventional ``estimate-then-quantize" scheme, under which the users independently estimate the correlated channels, is not optimal. In this paper, our goal is to leverage the correlation among different users' channels to reduce the overhead for CSI acquisition in IRS-assisted downlink communication. The main contributions of this paper are summarized as follows:
\begin{itemize}
    \item We consider an FDD IRS-assisted downlink communication system with multiple single-antenna users. Moreover, we assume that the channels are quasi-static block fading channel model and the direct channels between the BS and users are blocked. In the FDD IRS-assisted downlink communication system, the conventional ``estimate-then-quantize" scheme cannot utilize the channel correlation revealed in \cite{wang2020channel} because although the received signals of all the users are correlated, each user independently estimates its channels based on its own received signals. To overcome this issue, in this paper, we propose a novel ``quantize-then-estimate'' protocol. In sharp contrast to the ``estimate-then-quantize'' counterpart, our strategy makes each user first quantize its received pilot signals and then transmit the quantization bits to the BS. After de-quantization, the BS knows the pilot signals received by all the users, which contain the global CSI. Therefore, the BS can exploit the channel correlation revealed in \cite{wang2020channel} to jointly estimate the cascaded channels of all the users. The benefits of the proposed protocol for overhead reduction are two-fold. First, the BS is able to estimate the channels based on shorter pilot signals as shown in \cite{wang2020channel}, i.e., the number of time samples for BS's pilot transmission can be reduced. Second, because each user receives fewer pilot symbols, the number of quantization bits for feedback transmission is also reduced.
    \item Under our proposed protocol, we first design the codebook to quantize the received pilot signals via the Lloyd algorithm. Next, we design an efficient method such that the BS can estimate all the users' cascaded channels based on its received feedback from the users. Specifically, we select several reference users, and the BS should estimate the ratios among the power of channels of the non-reference users and that of the reference users based on the feedback received in the first a few time samples, and estimate the channels of the reference users based on the feedback received in the remaining time samples. The linear minimum mean-squared error (LMMSE) estimation technique is proposed to estimate the channel ratios and the channels under the above scheme. At last, we characterize the minimum overhead for transmitting the pilot signals in the downlink and the feedback signals in the uplink under our proposed ``quantize-then-estimate" protocol, and the significant overhead reduction compared to the conventional ``estimate-then-quantize" protocol is analytically demonstrated.
    \item To further improve the accuracy of CSI acquisition under our proposed protocol, we consider quantization bit allocation when users feed back their received pilot signals to the BS. Based on \cite{wang2022massive}, when the number of IRS sub-surfaces is large, the received pilot signals tend to be Gaussian distributed. Then, rely on the the approximate Gaussian test channel\cite{xia2006design}, we characterize the rate-distortion trade-off for feedback transmission and propose an efficient approach to design the quantization bit allocation of each user to optimize the rate-distortion trade-off.
\end{itemize}

\subsection{Organization}

The rest of this paper is organized as follows. Section \ref{sec:model} describes the system model for our considered IRS-assisted multi-user downlink communication system. Section \ref{sec:protocol} reviews the traditional ``estimate-then-quantize'' CSI acquisition protocol and introduces our proposed ``quantize-then-estimate'' protocol. Section \ref{sec:Design} describes how to implement the ``quantize-then-estimate" protocol in practice and characterize its minimum overhead. Section \ref{sec:rate distortion} designs a quantization bit allocation method for each user based on an approximated Gaussian test channel model. Section \ref{sec:numerical} provides numerical examples to demonstrate the effectiveness of our proposed ``quantize-then-estimate'' scheme. Section \ref{sec:conclusion} concludes the paper.

\textit{Notation}: $\bm I$ and $\bm O$ denote an identify matrix and an all-zero matrix, with appropriate dimensions. For a square full-rank matrix $\bm A$, ${\bm A}^{-1}$ denotes its inverse. For a matrix $\bm B$, ${\bm B}^{T}$, ${\bm B}^{H}$, and ${\bm B}^{\dagger}$ denotes its transpose, conjugate transpose, and pseudo-inverse matrix, respectively. $\lceil\cdot\rceil$ denotes the ceiling function. $\otimes$ denotes the Kronecker product.

\section{System Model}\label{sec:model}
\begin{figure}[t]
    \centering
    \includegraphics[scale=0.12]{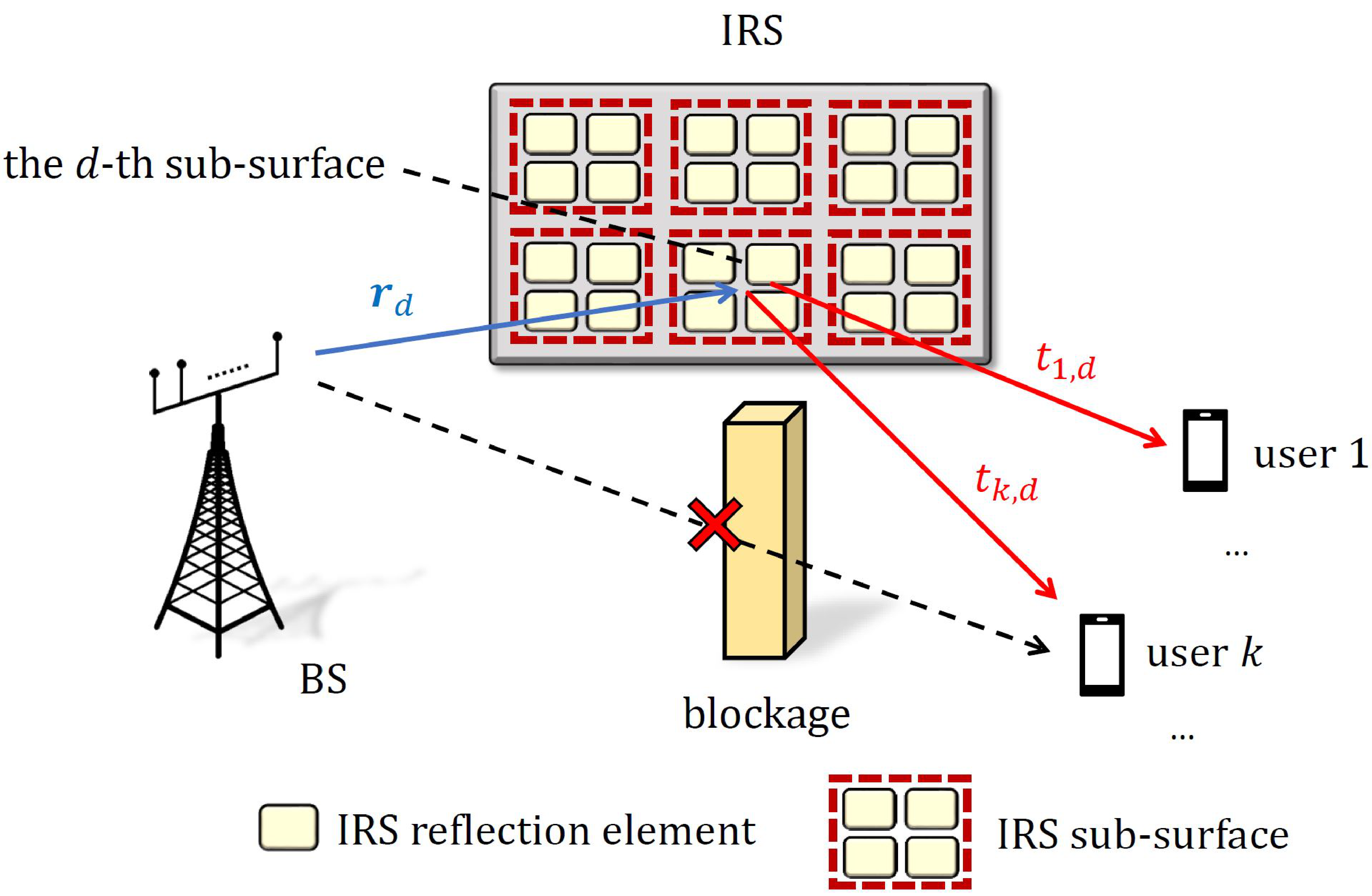}\caption{An IRS-aided downlink communication system.}
    \label{sys_model}
\end{figure}

We study the downlink communication in an FDD system which consists of a BS with $M$ antennas, $K$ single-antenna users, and an IRS with $N$ passive reflecting elements, as shown in Fig. \ref{sys_model}. In practice, the number of BS antennas is usually much larger than the number of users. Therefore, in this paper, we assume that $M>K$. In our considered IRS-assisted systems, the overhead to acquire CSI in IRS-assisted communication systems is high due to the large number of IRS elements $N$. To tackle this challenge, we adopt an IRS element grouping strategy as in \cite{yang2020intelligent,zheng2020intelligent}. Specifically, the IRS elements are divided into $D$ sub-surfaces to reduce the number of channels to be estimated, and the IRS elements within each sub-surface share a common reflection coefficient. Let ${\phi}_{d,i}=e^{j\theta_{d,i}}\in\mathbb{C}$ denote the reflection coefficient of the $d$-th IRS sub-surface at time sample $i$, where $\theta_{d,i}\in[0,2\pi)$ denotes the phase shift of $\phi_{d,i}$.

We assume a quasi-static block fading channel model, where the channels remain approximately constant in each coherence block. In addition, as shown in Fig. \ref{sys_model}, we assume that the direct channels between the BS and users are blocked, and the signals can only be transmitted through the IRS reflecting channels to the users. The baseband equivalent channels from the BS to the $d$-th IRS sub-surface, and from the $d$-th IRS sub-surface to user $k$ are denoted by ${\bm r}_{d}\in\mathbb{C}^{M\times 1}$ and ${t}_{k,d}\in\mathbb{C}, k=1,\cdots,K, d=1,\cdots,D$, respectively. For convenience, define ${\bm R}=[{\bm r}_1,\cdots,{\bm r}_{D}]$ as the overall channels from the BS to the IRS, and ${\bm t}_k=[{t}_{k,1},\cdots,{t}_{k,D}]^T$ as the channels from the IRS to user $k$. Then, the cascaded reflecting channels from the BS to user $k$ through the $d$-th IRS sub-surface is expressed as
\begin{align}\label{cas_ch}
    {\bm g}_{k,d}={t}_{k,d}{\bm r}_{d}\in\mathbb{C}^{M\times 1}, ~~\forall d, k.
\end{align}

In downlink communication, the pilot signals transmitted from the BS at time sample $i$ are denoted by ${\bm x}_i=[x_{i,1},\cdots,x_{i,M}]^T\in\mathbb{C}^{M\times1}, i=1,\cdots,T$, where $x_{i,m}$ is the $i$-th pilot sample transmitted by antenna $m$, and $T$ is the number of time samples to transmit the pilots. Then, the received signal of user $k$ at time sample $i$ is expressed as
\begin{align}\label{rev_com}
    y_{k,i} &= \sum_{d=1}^{D}{\phi}_{d,i}{\bm g}_{k,d}^T\sqrt{p_i}{\bm x}_i+z_{k,i},~~\forall k,i,
\end{align}
where $p_i$ denotes the transmit power at the BS at time sample $i$, and $z_{k,i}\sim\mathcal{CN}(0,\sigma_z^2)$ denotes the additive white Gaussian noise (AWGN) of user $k$ at time sample $i$.

\section{A Novel Quantize-then-Estimate Protocol}\label{sec:protocol}

In this paper, we aim to propose a low-overhead channel estimation and feedback scheme such that the BS can efficiently acquire the cascaded channels ${\bm g}_{k,d}$'s, $\forall k,d$, to design the BS beamforming vectors and the IRS reflecting coefficients. In this section, we will first overview the conventional method for channel estimation and feedback in our considered IRS-assisted FDD systems, and show its disadvantages. Then, we will propose a novel strategy that can exploit the unique channel property in IRS-assisted communication to reduce the channel estimation and feedback overhead.

\subsection{Traditional Estimate-then-Quantize Scheme}\label{subsec:conv}

Without IRS, channel estimation and feedback have been widely studied for downlink FDD systems \cite{jindal2006mimo,rao2014distributed,love2008overview}.  In the IRS-assisted network, we may adopt similar philosophy and apply the following ``estimate-then-quantize" strategy for channel estimation and feedback.
\begin{itemize}
    \item \textbf{Phase I (Estimation)}: In the first phase, each user $k$ first receives the pilot signals from the BS as shown in \eqref{rev_com} and then estimates its cascaded channel ${\bm g}_{k,d}$'s, $\forall d$, based on the existing method proposed in
    \cite{yang2020intelligent,zheng2020intelligent,mishra2019channel,alwazani2020intelligent}.
    \item \textbf{Phase II (Feedback)}: In the second phase, each user quantizes its estimated channels and feeds back the quantization bits to the BS. The BS de-quantizes the quantization bits to recover the cascaded channels of all the users.
\end{itemize}

Although the above ``estimate-then-quantize'' strategy works well in conventional systems without IRS, its overall overhead for pilot and feedback transmission is significant in our considered IRS-assisted systems. Specifically, under the above scheme, the minimum number of pilot samples for user $k$ to estimate ${\bm g}_{k,d}$'s, $d=1,\cdots,D$, in Phase I is $T_{\rm min}=MD$ \cite{wang2020channel}. Moreover, each user $k$ needs to feed back $MD$ channel coefficients in ${\bm g}_{k,d}$'s, $d=1,\cdots,D$, to the BS in Phase II. Recently, it has been revealed in \cite{wang2020channel} that there is a lot of redundancy in users' cascaded channels and the number of independent unknown variables in ${\bm g}_{k,d}$'s, $k=1,\cdots,K, d=1,\cdots,D$, is much smaller than $KMD$. Specifically, if we focus on a particular IRS sub-surface $d$, then the channel between the BS and the $d$-th sub-surface of the IRS, i.e., ${\bm r}_d$, is common among the cascaded channels ${\bm g}_{k,d}$'s of all the users. As a result, according to \eqref{cas_ch}, we have
\begin{align}\label{corr_group}
    {\bm g}_{k,d}=\lambda_{k,d}{\bm g}_{k_d,d}, ~~\forall k\neq k_d, ~d=1,\cdots,D,
\end{align}
where $k_d$ is the index of the reference user selected for IRS sub-surface $d$, and the channel ratio between user $k$ and reference user $k_d$ is given by
\begin{align}\label{ratio}
    \lambda_{k,d}=\frac{{t}_{k,d}}{{t}_{k_d,d}}.
\end{align}
\eqref{corr_group} and \eqref{ratio} indicate that for each IRS sub-surface $d$, if the cascaded channel of the reference user $k_d$, i.e., ${\bm g}_{k_d,d}$, is known, a scalar $\lambda_{k,d}$ is sufficient for the BS to know the cascaded channel vector of user $k\neq k_d$, i.e., ${\bm g}_{k,d}$. In other words, the BS just needs to know $MD+(K-1)D$ channel coefficients in ${\bm g}_{k_d,d}$'s, $d=1,\cdots,D$, and $\lambda_{k,d}$'s, $\forall k\neq k_d$, $d=1,\cdots,D$. Therefore, the number of time samples for pilot transmission in Phase I and the number of quantization bits in Phase II can be hugely reduced in IRS-assisted downlink communication, if the channel property shown in \eqref{corr_group} and \eqref{ratio} can be properly utilized. However, the conventional ``estimate-then-quantize'' scheme does not take advantage of \eqref{corr_group} and \eqref{ratio} for reducing the overhead.

\subsection{Proposed Quantize-then-Estimate Scheme}\label{subsec:proposed}

In this sub-section, we propose a novel strategy that can leverage \eqref{corr_group} and \eqref{ratio} to significantly reduce the overhead for channel estimation and feedback in IRS-assisted downlink communication. Note that \eqref{corr_group} and \eqref{ratio} reveal the correlation among different users' cascaded channels, while the users are distributed and cannot cooperate with each other to leverage such correlation for channel estimation. To overcome this issue, we propose a ``quantize-then-estimate'' protocol, where the channels of all the users are estimated at the BS side by utilizing \eqref{corr_group} and \eqref{ratio}, rather than at the distributed user side. The proposed scheme is detailed as below.

\begin{itemize}
    \item \textbf{Phase I (Feedback)}: In the first phase, all the users receive the pilot signals from the BS as shown in \eqref{rev_com}. However, instead of estimating its own channels, each user $k$ quantizes its received pilot signals over $T$ time samples, i.e., ${\bm y}_k=[y_{k,1},...,y_{k,T}]^T$, and feeds back the quantization bits to the BS.
    \item \textbf{Phase II (Estimation)}: In the second phase, the BS de-quantizes the received quantization bits for recovering the pilot signals received by all the users. Then, with the above global information about users' received pilots, the BS is able to leverage the correlation among different users' channels shown in \eqref{corr_group} and \eqref{ratio} to estimate ${\bm g}_{k,d}$'s more efficiently.
\end{itemize}

Note that the key difference between our proposed ``quantize-then-estimate'' scheme and the conventional ``estimate-then-quantize'' scheme shown in Section \ref{subsec:conv} lies in what is fed back from the users to the BS and who performs channel estimation. Under our scheme, users feed back their received pilot signals such that the BS can perform a joint estimation of different users' channels by leveraging \eqref{corr_group} and \eqref{ratio}. The benefits of the above joint estimation are two-fold. First, as shown in \cite{wang2020channel}, the minimum number of pilot samples for channel estimation, i.e., $T_{\rm min}$, can be reduced from $MD$ to $\max\{M-1, D+\lceil (M-1)D/K \rceil\}$ by leveraging \eqref{corr_group} and \eqref{ratio}. Second, because signals from fewer time samples are quantized, the feedback overhead is significantly reduced. Specifically, without utilizing \eqref{corr_group} and \eqref{ratio}, all the users need to quantize $KMD$ samples in ${\bm g}_{k,d}$'s; while under our proposed scheme, as will be shown later in Section \ref{sec:Design}, all the users only need to quantize $KD+K\cdot\max\{M-1, D+\lceil (M-1)D/K \rceil\}$ samples in ${\bm y}_k$'s.

\section{Quantization and Estimation Design}\label{sec:Design}

In this section, we will introduce how to quantize $y_{k,i}$'s at the user side and how to estimate the channels via leveraging \eqref{corr_group} and \eqref{ratio} at the BS side, in Phase I and Phase II of our proposed ``quantize-then-estimate'' protocol, respectively. We will also analytically characterize the overhead for pilot and feedback transmission under our proposed scheme and show the significant overhead reduction over the conventional ``estimate-then-quantize'' protocol.

\subsection{Quantization at User Side}\label{subsec:quan}

In this sub-section, we introduce how the users quantize the received pilot signals. For each user $k$, we assume independent signal quantization at different time samples. Specifically, each user $k$ quantizes ${y}_{k,1},\cdots,{y}_{k,T}$ subsequently via scalar quantization. At time sample $i$, the codebook for quantizing ${y}_{k,i}$ is denoted by
\begin{align}\label{codebook}
    \mathcal{C}_{k,i}=\{{c}_{k,i,1},\cdots,{c}_{k,i,L_{k,i}}\},~~\forall k,i,
\end{align}
which consists of $L_{k,i}$ codewords and is shared by user $k$ and the BS. We will introduce how to design $L_{k,i}$'s, $\forall k$ and $i$, in Section \ref{sec:rate distortion}. Based on the probability density function of ${y}_{k,i}$, the codebook given any $L_{k,i}$ can be designed via the Lloyd algorithm \cite{gersho2012vector}. Given the codebook $\mathcal{C}_{k,i}$, the codeword index and the corresponding codeword to quantize ${y}_{k,i}$ are given by
\begin{align}
    &l_{k,i}^{\ast}=\underset{l_{k,i}=1_{k,i},\cdots,L_{k,i}}{\arg\min}~\mathcal{D}({y}_{k,i},{c}_{k,i,l}), \notag\\ &\tilde{y}_{k,i}=c_{_{k,i,l^{\ast}_{k,i}}},~~\forall k,i,
\end{align}
where $\mathcal{D}({y}_{k,i},{c}_{k,i,l})=\|{y}_{k,i}-{c}_{k,i,l}\|^2_2$ denotes the distortion function between ${y}_{k,i}$ and the codeword $c_{k,i,l}$. The quantized signal of ${y}_{k,i}$ can be expressed as
\begin{align}\label{yki_quan}
    \tilde{y}_{k,i}={y}_{k,i}+{e}_{k,i},~~\forall k,i,
\end{align}
where $e_{k,i}$ denotes the error to quantize the signal ${y}_{k,i}$ with zero mean and variance $q_{k,i}$. Note that $e_{k,i}$'s are independent over $i$ due to scalar quantization at each time sample.

For user $k$, each codeword index $l_{k,i}^{\ast}$ is represented by $\lceil\log_2 L_{k,i}\rceil$ quantization bits. Since independent quantization is performed at
different time samples, the total number of bits for user $k$ to quantize all its received pilot signals is the sum of the number of bits over $T$ time samples, i.e.,
\begin{align}
    B_k=\sum_{i=1}^T\lceil\log_2 L_{k,i}\rceil, ~~\forall k.
\end{align}
Then, each user modulates its quantization bits onto quadrature amplitude modulation (QAM) symbols and sends these symbols to the BS via a feedback channel, which is assumed to be error-free \cite{caire2010multiuser}. Denote the modulation rate of user $k$ as $\mu_k$ bits/sample, $\forall k$. Thus, the number of time samples for user $k$ to feed back the QAM symbols to the BS is
\begin{align}
    T_{{\rm fb},k}=\frac{B_k}{\mu_k}=\frac{1}{\mu_k}\sum_{i=1}^T\lceil\log_2 L_{k,i} \rceil, ~~\forall k.
\end{align}
As show in \cite{caire2010multiuser}, in our considered setup where the number of BS antennas is larger than that of users, i.e., $M>K$, the BS can mitigate inter-user interference via zero-forcing beamforming design. Therefore, all the users can transmit the feedback symbols simultaneously to the BS  without inter-user interference. As a result, the number of time samples required for all the users to finish feedback transmission is determined by the user that needs the largest number of time samples, i.e.,
\begin{align}\label{T_fb}
    T_{\rm fb}=\underset{1\leq k\leq K}{\max}~ T_{{\rm fb},k}.
\end{align}

\subsection{Channel Estimation at BS Side}\label{subsec:ch_es}

In this sub-section, we introduce how the BS can estimate the channels. To begin with, the BS collects the QAM symbols from the users and de-modulates the symbols to quantization bits. Similar to \cite{caire2010multiuser}, we assume that the feedback channels from the users to the BS are error-free such that the BS can perfectly decode the quantization bits and then recover $\tilde{y}_{k,i}$'s.

Subsequently, based on the de-quantized signals $\tilde{y}_{k,i}$'s, the BS adopts a two-step channel estimation method, where in the first step with a duration of $\tau_1<T$ samples, the BS estimates $\lambda_{k,d}$'s, $\forall k\neq k_d$ and $\forall d$, based on  $\tilde{y}_{k,1},\cdots,\tilde{ y}_{k,\tau_1}$ for non-reference users;
while in the second step with a duration of $\tau_2=T-\tau_1$ samples, the BS estimates ${\bm g}_{k_d,d}$'s based on $\tilde{y}_{k,\tau_1+1},\cdots,\tilde{y}_{k,\tau_1+\tau_2}$, $\forall k,d$, for reference users. Last, user $k$'s cascaded channels can be recovered based on \eqref{corr_group}, $\forall k$. In the following, we introduce how to estimate $\lambda_{k,d}$'s in Step 1 and ${\bm g}_{k_d,d}$'s in Step 2.

\textbf{Step 1} (\textit{Estimation of channel ratios $\lambda_{k,d}$'s}): In the first step, the de-quantized received signals $\tilde{y}_{k,i}$'s given in \eqref{yki_quan} can be re-written as
\begin{align}\label{yki_quan_1}
    \tilde{y}_{k,i}=\sum_{d=1}^{D}{\phi}_{d,i}\alpha_{k,d,i}+z_{k,i}+e_{k,i},~~\forall k, i=1,\cdots,\tau_1,
\end{align}
where
\begin{align}\label{alpha_i}
    \alpha_{k,d,i}=\begin{cases}
        $$ \sqrt{p_i}{\bm g}_{k_d,d}^T{\bm x}_i
         $$, &\mbox{if $k=k_{d}$}, \\
        $$  \sqrt{p_i}{{\lambda}}_{k,d}{\bm g}_{k_d,d}^T{\bm x}_i $$, &\mbox{if $k\neq k_{d}$},
    \end{cases} ~~\forall d,i=1,\cdots,\tau_1.
\end{align}
Note that if we can perfectly recover $\alpha_{k,d,i}$'s in \eqref{alpha_i}, then the channel ratio can be obtained by
\begin{align}\label{corr_es_kd}     {\lambda}_{k,d}=\frac{\alpha_{k,d,i}}{\alpha_{k_d,d,i}}, ~~\forall k\neq k_d,\forall d, i=1,\cdots,\tau_1.
\end{align}
The main challenge to estimate $\alpha_{k,d,i}$'s in \eqref{alpha_i} is that for each $k$, the BS has to estimate $\tau_1D$ unknown variables, i.e., $\alpha_{k,d,i}$'s, $d=1,\cdots,D, i=1,\cdots,\tau_1$, using $\tau_1<\tau_1D$ observations, i.e., $\tilde{y}_{k,i}$'s, $i=1,\cdots,\tau_1$.
To solve this challenge, we propose to set identical pilot signals over all time samples, i.e.,
\begin{align}
    \sqrt{p_i}{\bm x}_i=\sqrt{p}{\bm x}, ~~\forall i.
\end{align}
In this case, the received signal in \eqref{yki_quan_1} reduces to
\begin{align}\label{yki_alpha}
    \tilde{y}_{k,i}=\sum_{d=1}^{D}{\phi}_{d,i}{\alpha}_{k,d}+z_{k,i}+e_{k,i}, ~~\forall k, i=1,\cdots,\tau_1,
\end{align}
where
\begin{align}\label{alpha}
    {\alpha}_{k,d}=
    \begin{cases}
        $$ \sqrt{p}{\bm g}_{k_d,d}^T{\bm x}$$, &\mbox{if $k=k_{d}$},\\
        $$ \sqrt{p}{\lambda}_{k,d}{\bm g}_{k_d,d}^T{\bm x} $$, &\mbox{if $k\neq k_{d}$},
    \end{cases}~~\forall d.
\end{align}
Note that in \eqref{yki_alpha}, for each $k$, there are $D$ unknown variables, i.e., $\alpha_{k,d}$, $d=1,\cdots,D$, rather than $\tau_1D$ variables as in \eqref{alpha_i}. Moreover, if $\alpha_{k,d}$'s can be perfectly estimated, $\lambda_{k,d}$'s can be estimated as
\begin{align}
    \lambda_{k,d}=\frac{\alpha_{k,d}}{\alpha_{k_d,d}},~~\forall k\neq k_d,\forall d.
\end{align}
In the following, we show how to estimate $\alpha_{k,d}$'s based on \eqref{yki_alpha}. Let $\tilde{\bm y}_k^{(1)}=[\tilde{y}_{k,1},\cdots,\tilde{y}_{k,\tau_1}]^T$ denote the overall quantized received signals of user $k$ over $\tau_1$ time samples. Then, \eqref{yki_alpha} is equivalent to
\begin{align}\label{rev1q}
    \tilde{\bm y}_{k}^{(1)}={\bm \Phi}_1{\bm \alpha}_{k}+{\bm z}_{k}^{(1)}+{\bm e}_{k}^{(1)},~~\forall k,
\end{align}
where
\begin{align}
    {\bm \Phi}_1=\left[
    \begin{array}{ccc}
        {\phi}_{1,1} & \cdots & {\phi}_{D,1}  \\
        \vdots & \ddots & \vdots \\
        {\phi}_{1,\tau_{1}} & \cdots & {\phi}_{D,\tau_{1}}
    \end{array} \right],
\end{align}
${\bm\alpha}_k=[\alpha_{k,1},\cdots,\alpha_{k,D}]^T$, ${\bm z}_k^{(1)}=[z_{k,1},\cdots,z_{k,\tau_1}]^T$,
and ${\bm e}_{k}^{(1)}=[e_{k,1},\cdots,e_{k,\tau_1}]^T$. If there is no noise and quantization error, i.e., ${\bm z}_k^{(1)}={\bm 0}$ and ${\bm e}_{k}^{(1)}={\bm 0}$, then we can set ${\bm \Phi}_1$ as a discrete Fourier transform (DFT) matrix and then perfectly estimate ${\bm \alpha}_k$ using
\begin{align}\label{tau1}
    \bar{\tau}_{1}=D
\end{align}
time samples. In the practical case with noise and quantization error, we can set $\tau_1\geq D$ and ${\bm \Phi}_1$ as the first $D$ columns of a $\tau_1\times \tau_1$ DFT matrix. Then, we can apply the LMMSE technique to estimate ${\bm \alpha}_k$ as
\begin{align}\label{es_alpha}
    \hat{\bm \alpha}_{k}
    &=[\hat{\alpha}_{k,1},\cdots,\hat{\alpha}_{k,D}]^T\notag\\
    &=
    {\bm \Lambda}_{k}{\bm \Phi}_1^H\left(
    {\bm \Phi}_1{\bm \Lambda}_{k}{\bm \Phi}_1^H+\sigma_z^2{\bm I}_D+{\bm E}_{k}^{(1)}
    \right)^{-1}\tilde{\bm y}_{k}^{(1)},
\end{align}
where ${\bm \Lambda}_{k}$ denotes the covariance matrix of ${\bm \alpha}_{k}$, and ${\bm E}_{k}^{(1)}$ denotes the covariance matrix of ${\bm e}_k^{(1)}$. Then, the estimated ${\lambda}_{k,d}$ is expressed as
\begin{align}\label{es_lambda}
    \hat{\lambda}_{k,d}=\frac{\hat{\alpha}_{k,d}}{\hat{\alpha}_{k_d,d}},~~\forall k\neq k_d,\forall d.
\end{align}
Given any reference user selection strategy $k_d$'s, $d=1,\cdots,\\ D$, we can estimate $\lambda_{k,d}$'s, $k\neq k_d$, using the above method. However, the selection of the reference user for each IRS sub-surface $d$ can significantly affect the accuracy for estimating $\lambda_{k,d}$'s, $\forall k\neq k_d, \forall d$. This is because if user $k$ with a very weak value of $|\hat{\alpha}_{k,d}|$ is selected as the reference user, then a very small error for estimating $\alpha_{k,d}$, i.e., $\hat{\alpha}_{k,d}-\alpha_{k,d}$, can cause a significant error for estimating $\lambda_{k,d}$ in \eqref{es_lambda}. Therefore, for each IRS sub-surface $d$, we select the reference user as follows
\begin{align}\label{k_d}
    k_{d}=\underset{k=1,\cdots,K}{\arg\max}\left|\hat{\alpha}_{k,d}\right|^2,~~\forall d.
\end{align}

\textbf{Step 2} (\textit{Estimation of reference users' channels ${\bm g}_{k_d,d}$'s}): In the second step, we estimate the channels of reference users, i.e., ${\bm g}_{k_d,d}$'s, $d=1,\cdots,D$. Before introducing Step 2, we want to emphasize that after $\hat{\alpha}_{k,d}$'s are estimated in Step 1, we already have some useful information about ${\bm g}_{k_d,d}$'s according to \eqref{alpha}:
\begin{align}
    \hat{\alpha}_{k_d,d}=\sqrt{p}{\bm x}^T{\bm g}_{k_d,d}+\beta_{k_d,d}, ~~d=1,\cdots,D,
\end{align}
where $\beta_{k_d,d}$ denotes the error for estimating ${\alpha}_{k_d,d}$. Define that $\hat{\bm \alpha}=[\hat{\alpha}_{k_1,1},\cdots,\hat{\alpha}_{k_D,D}]^T$ and ${\bm g}=[{\bm g}_{k_1,1}^T,\cdots,{\bm g}_{k_D,D}^T]^T$. Then, we have
\begin{align}\label{es_alpha_kd}
    \hat{\bm \alpha}={\bm F}{\bm g}+{\bm \beta},
\end{align}
where ${\bm F}=\sqrt{p}{\bm I}_D\otimes{\bm x}^T\in\mathbb{C}^{D\times MD}$, and ${\bm \beta}=[{\beta}_{k_1,1},\cdots,{\beta}_{k_D,D}]^T$. This information should be used in Step 2 to estimate ${\bm g}$.

In Step 2, the received pilots at time sample $i$ is given as
\begin{align}\label{rev2q}
    \tilde{\bm y}_{i}^{(2)}&=
    [\tilde{y}_{1,i},\cdots,\tilde{y}_{K,i}]^T=\sqrt{p}\sum_{d=1}^{D}{\phi}_{d,i}{\bm x}_{i}^T{\bm g}_{k_d,d}{\bm \lambda}_d+{\bm e}_i^{(2)} \notag\\
    &={\bm F}_i{\bm g}+{\bm e}_i^{(2)},~~i=\tau_1+1,\cdots,\tau_1+\tau_2,
\end{align}
where ${\bm \lambda}_{d}=[{\lambda}_{1,d},\cdots,{\lambda}_{K,d}]^T$ with ${\lambda}_{k_d,d}=1$, $\forall d$, ${\bm e}_i^{(2)}=[z_{1,i}+e_{1,i},\cdots,z_{K,i}+e_{K,i}]^T$, and
\begin{align}\label{Fi}
    {\bm F}_i=\sqrt{p}[\phi_{1,i}{\bm x}_i^T\otimes{\bm \lambda}_1,\cdots,\phi_{D,i}{\bm x}_i^T\otimes{\bm \lambda}_D]\in\mathbb{C}^{K\times MD}.
\end{align}
Since both $\hat{\bm \alpha}$ in Step 1 and $\tilde{\bm y}_{i}^{(2)}$'s in Step 2 contain information about ${\bm g}$, we define
\begin{align}
   \tilde{\bm y}^{(2)}=\left[(\tilde{\bm y}_{\tau_1+1}^{(2)})^T,\cdots,(\tilde{\bm y}_{\tau_1+\tau_2}^{(2)})^T, \hat{\bm\alpha}^T\right]^T.
\end{align}
According to \eqref{es_alpha_kd} and \eqref{rev2q}, we have
\begin{align}\label{ystep2}
    \tilde{\bm y}^{(2)}&={\bm\Theta}{\bm g}+{\bm e}^{(2)}\notag\\
    &=\hat{\bm\Theta}{\bm g}+({\bm \Theta}-\hat{\bm\Theta}){\bm g}+{\bm e}^{(2)},
\end{align}
where ${\bm \Theta}=\left[{\bm F}_{\tau_1+1}^T,\cdots,{\bm F}_{\tau_1+\tau_2}^T,{\bm F}^T \right]^T$, $\hat{\bm\Theta}$ is an estimation of $\bm \Theta$ with $\lambda_{k,d}$'s (as shown in \eqref{Fi}, ${\bm F}_i$'s are functions of $\lambda_{k,d}$'s) replaced by their estimations $\hat{\lambda}_{k,d}$'s given in \eqref{es_lambda}, and ${\bm e}^{(2)}=\left[({\bm e}_{\tau_1+1}^{(2)})^T,\cdots,({\bm e}_{\tau_1+\tau_2}^{(2)})^T,{\bm \beta}^T\right]^T$. Note that in \eqref{ystep2}, $\hat{\bm \Theta}$ is known by the BS, and ${\bm\Theta}-\hat{\bm\Theta}$ is the unknown estimation error. If there is no noise and quantization error in both Step 1 (such that \
$\hat{\lambda}_{k,d}=\lambda_{k,d}$, $\forall k,d$, and $\hat{\bm \Theta}={\bm \Theta}$) and Step 2, i.e., ${\bm e}^{(2)}={\bm 0}$, then \eqref{ystep2} reduces to
\begin{align}\label{tilde_y2}
    \tilde{\bm y}^{(2)}={\bm\Theta}{\bm g}.
\end{align}
\vspace{-15pt}
\begin{theorem}\label{theorem1}
    In the ideal case that there is no noise in users' received pilots given in \eqref{rev_com} and quantization error in BS's received signals given in \eqref{ystep2}, i.e., ${\bm \Theta}=\hat{\bm \Theta}$ and ${\bm e}^{(2)}={\bm 0}$, the minimum number of time samples for the BS to perfectly estimate $\bm g$ based on \eqref{tilde_y2} is
    \begin{align}\label{tau2}
        \bar{\tau}_{2}=\max\left\{M-1, \left\lceil\frac{(M-1)D}{K}\right\rceil\right\}.
    \end{align}
\end{theorem}
\begin{IEEEproof}
    Please refer to Appendix.
\end{IEEEproof}

To summarize, the minimum number of time samples to transmit pilot signals from the BS to the users is
\begin{align}\label{T_es_min}
    T_{\rm min}^{\ast}=\bar{\tau}_1+\bar{\tau}_2=D+\max\left\{M-1, \left\lceil\frac{(M-1)D}{K}\right\rceil \right\}.
\end{align}

In the practical case with noise and quantization error, we can use $\tau_2\geq\bar{\tau}_2$ time samples for pilot transmission. However, the unknown error propagated from Step 1, i.e., ${\bm \Theta}-\hat{\bm \Theta}$ in \eqref{ystep2}, makes it hard to obtain the LMMSE estimator of $\bm g$. Actually, we can increase the number of time samples for pilot transmission in Step 1 such that ${\bm \Theta}-\hat{\bm \Theta}$ is sufficiently small. In this case, we assume that ${\bm \Theta}-\hat{\bm \Theta}\approx {\bm 0}$ as in \cite{wang2021new}. Then, \eqref{ystep2} reduces to
\begin{align}\label{y2q_approx}
    \tilde{\bm y}^{(2)}\approx\hat{\bm\Theta}{\bm g}+{\bm e}^{(2)}.
\end{align}
Based on \eqref{y2q_approx}, the LMMSE estimator of $\bm g$ can be designed. Specifically, we can set pilot signals ${\bm x}_i$'s and IRS reflection coefficients $\phi_{d,i}$'s, $d=1,\cdots,D$, $i=\tau_1+1,\cdots,\tau_1+\tau_2$, according to the orthogonal transmission
and reflection strategy in Section V-C in \cite{wang2020channel}. In this case, based on \eqref{y2q_approx}, the LMMSE estimator of $\bm g$ is given as
\begin{align}\label{es_g}
    \hat{\bm g}
    ={\bm G}\hat{\bm \Theta}^H\left(
    \hat{\bm \Theta}{\bm G}\hat{\bm \Theta}^H+{\bm E}^{(2)}
    \right)^{-1}\tilde{\bm y}^{(2)},
\end{align}
where ${\bm G}$ denotes the covariance matrix of $\bm g$, and ${\bm E}^{(2)}$ denotes the covariance matrix of ${\bm e}^{(2)}$. With the estimations of $\lambda_{k,d}$'s given in \eqref{es_lambda} and those of ${\bm g}_{k_d,d}$'s given in \eqref{es_g}, the cascaded channels can be estimated as
\begin{align}
    \hat{\bm g}_{k,d}=\hat{\lambda}_{k,d}\hat{\bm g}_{k_d,d},~~k=1,\cdots,K,d=1,\cdots,D.
\end{align}

\begin{remark}
Until now, we have shown that under our proposed quantize-and-estimate protocol, the minimum numbers of time samples for pilot transmission and feedback transmission are \eqref{T_es_min} and \eqref{T_fb}, respectively. Therefore, the minimum number of time samples for pilot and feedback transmission is
\begin{align}\label{T_tol}
    T_{\rm tol}&=T_{\rm min}^{\ast}+T_{\rm fb} \notag\\
    &=T_{\rm min}^{\ast}+\underset{1\leq k\leq K}{\max}\left\{\frac{1}{\mu_k}\sum_{i=1}^{T_{\rm min}^{\ast}}\lceil\log_2 L_{k,i}\rceil\right\}.
\end{align}
Note that under the conventional ``estimate-then-quantize'' strategy, the minimum number of time samples for pilot transmission is \cite{wang2020channel}
\begin{align}\label{T_es_min_cov}
    T_{\rm cov}^{\ast}=MD.
\end{align}
Then, the minimum number of time samples for pilot and feedback transmission is
\begin{align}\label{T_tol_cov}
    T_{\rm tol, cov}=T_{\rm cov}^{\ast}+\underset{1\leq k\leq K}{\max}\left\{\frac{MD\lceil\log_2 L_{k}\rceil}{\mu_k}\right\},
\end{align}
where $L_k$ is the codebook size for each user $k$ to quantize each estimated channel coefficient. Because fewer pilot samples are transmitted in Phase I and quantized in Phase II thanks to the utilization of channel correlation shown in \eqref{corr_group}, the overhead of our proposed ``quantize-then-estimate'' strategy characterized in \eqref{T_tol} is significantly reduced compared with that of the conventional ``estimate-then-quantize'' strategy characterized in \eqref{T_tol_cov}.
\end{remark}

At last, given the minimum number of time samples for pilot and
feedback transmission, we characterize the computational complexity of our proposed scheme and the conventional ``estimate-then-quantize'' scheme. For the feedback phase, we count the total time of exhaustively searching the codebook for selecting the optimal codeword as the measure of complexity. If $B$ bits are used to quantize each complex symbol, then the complexity in the feedback phase under our proposed scheme is $\mathcal{O}((D+\lceil(M-1)D/K\rceil)2^B)$, and that under the ``estimate-then-quantize'' scheme is $\mathcal{O}(MD2^B)$, respectively. For the estimation phase, we count the number of complex multiplication (CM) \cite{chen2023channel} as the measure of complexity. The main complexity of our proposed scheme comes from computing \eqref{es_alpha} and \eqref{es_g}, It is straightforward
to see that the two operations require $\Delta_1=D^2(1+\beta+3D)$ CMs and $\Delta_2=(K\lceil(M-1)D/K\rceil+D)(MD)^2+(K\lceil(M-1)D/K\rceil+D)^2(2MD+\beta+1)$ CMs, respectively, where $\beta$ is a scaling factor depending on the specific algorithm for the matrix inversion. As a result, the total computational complexity can be expressed as $\mathcal{O}(K\Delta_1+\Delta_2)$. The complexity of the ``estimate-then-quantize'' scheme can be expressed as $\mathcal{O}((MD)^2(3MD+\beta+1))$. Therefore, the complexity in the estimation phase of the two schemes are both nearly $\mathcal{O}((MD)^3)$, while the complexity in the feedback phase under our proposed scheme is much lower than that under the ``estimate-then-quantize'' scheme.

\section{Quantization Bit Allocation}\label{sec:rate distortion}

In section \ref{sec:Design}, we have proposed efficient methods to quantize pilot signals at the user side and estimate the channels at the BS side. A remaining issue that is not tackled is how to determine the quantization bit allocation policy for each user to achieve the best rate-distortion trade-off under our proposed ``quantize-then-estimate'' scheme. The quantized version of user $k$'s received pilot signal at time sample $i$ is given in \eqref{yki_quan}. However, the distribution of the quantization errors under the Lloyd algorithm is usually non-trivial, and it is thus difficult to analyze the rate-distortion trade-off. In this section, we will apply the Gaussian test channel model to approximate the rate-distortion trade-off achieved by the Lloyd algorithm, based on which we are able to optimize the quantization bit allocation, i.e., $L_{k,i}$'s. As will be shown later in this section, the Gaussian test channel model yields an analytical expression for the rate-distortion performance. Moreover, \cite{xia2006design} has shown analytically and numerically that the rate-distortion trade-off obtained under the Gaussian test channel model is a very tight approximation to that achieved by the practical Lloyd algorithm. We will also numerically verify the tightness of the Gaussian test channel model in terms of rate-distortion approximation later in this section.

\subsection{Gaussian Test Channel and Rate-Distortion Trade-off}\label{subsec:testc}

The Gaussian test channel model to approximate \eqref{yki_quan} is given as
\begin{align}\label{test}
    \tilde{y}_{k,i}={y}_{k,i}+\tilde{e}_{k,i},~~\forall k,i,
\end{align}
where $\tilde{e}_{k,i}$ has the same variance as ${e}_{k,i}$ in \eqref{yki_quan} and is assumed to be Gaussian distributed, i.e., $\tilde{e}_{k,i}\sim\mathcal{CN}(0,q_{k,i})$, and is independent with ${y}_{k,i}$, $\forall k,i$. Moreover, because each user $k$ applies scalar quantization on $y_{k,i}$'s, $\forall i$, $\tilde{e}_{k,i}$'s are independent over $i$.

As shown in Lemma $1$ of \cite{wang2022massive}, when the number of IRS sub-surfaces $D$ is large, $y_{k,i}$, $\forall k,i$, tends to be Gaussian distributed in general, i.e., $y_{k,i}\sim\mathcal{CN}(0,u_{k,i})$, where $u_{k,i}=\mathbb{E}[|y_{k,i}|^2]$ denotes the variance of $y_{k,i}$, $\forall k,i$. Then, under the Gaussian test channel model given in \eqref{test}, the BS can adopt an MMSE estimator to recover ${y}_{k,i}$ based on $\tilde{y}_{k,i}$, and the corresponding estimation MSE is \cite{Pishro2014probability}
\begin{align}
    \gamma_{k,i}
    =\frac{u_{k,i}q_{k,i}}{u_{k,i}+q_{k,i}}~~\forall k,i.
\end{align}
Therefore, given ${\bm q}_{k}=[q_{k,1},\cdots,q_{k,T}]^T$, the overall MSE to estimate ${\bm y}_{k}=[y_{k,1},\cdots,y_{k,T}]^T$ is given as
\begin{align}\label{distortion}
    \Gamma_k({\bm q}_k)=\sum_{i=1}^T\gamma_{k,i}
    =\sum_{i=1}^{T}
    \frac{u_{k,i}q_{k,i}}{u_{k,i}+q_{k,i}},~~\forall k.
\end{align}
Last, because different users independently quantize their received signals, the overall MSE to estimate ${\bm y}_1,\cdots,{\bm y}_K$ is given as
\begin{align}\label{distortion_all}
    \Gamma_{\rm sum}({\bm q})=\sum_{k=1}^K\Gamma_k({\bm q}_k)=\sum_{k=1}^K\sum_{i=1}^T\frac{u_{k,i}q_{k,i}}{u_{k,i}+q_{k,i}},
\end{align}
where ${\bm q}=[{\bm q}_1^T,\cdots,{\bm
 q}_K^T]^T$. Moreover, according to \eqref{test}, the number of bits to quantize the sample $y_{k,i}$ is \cite{el2011network}
\begin{align}\label{bits_ki}
    \mathcal{I}(\tilde{y}_{k,i},{y}_{k,i})&=\mathcal{H}(\tilde{y}_{k,i})-\mathcal{H}(\tilde{y}_{k,i}|y_{k,i})=\mathcal{H}(\tilde{y}_{k,i})-\mathcal{H}(\tilde{e}_{k,i}) \notag\\
    &=\log_2\left[\pi e (u_{k,i}+q_{k,i})\right]-\log_2(\pi e q_{k,i}) \notag\\
    &=\log_2 \left(1+\frac{u_{k,i}}{q_{k,i}}\right), ~~\forall k,i,
\end{align}
where $\mathcal{I}(\tilde{y}_{k,i},{y}_{k,i})$ is the mutual information between $\tilde{y}_{k,i}$ and $y_{k,i}$, $\mathcal{H}(\cdot)$ denotes the differential entropy, and the second equality is due to independence of $y_{k,i}$ and $\tilde{e}_{k,i}$. Then, the overall number of bits for user $k$ to quantize ${\bm y}_k$ over $T$ time samples is
\begin{align}\label{B_k_G}
    B_k^{(G)}({\bm q}_k)=\sum_{i=1}^T\log_2 \left(1+\frac{u_{k,i}}{q_{k,i}}\right), ~~\forall k.
\end{align}
Thus, similar to $\eqref{T_fb}$, the feedback transmission time (in terms of samples) under the Gaussian test channel model \eqref{test} is
\begin{align}\label{T_fb_G}
    T_{\rm fb}^{(G)}=\underset{1\leq k \leq K}{\max}~~T_{{\rm fb},k}^{(G)},
\end{align}
where
\begin{align}\label{T_fb_k_G}
    T_{{\rm fb},k}^{(G)}=\frac{B_k^{(G)}({\bm q}_k)}{\mu_k},~~\forall k.
\end{align}

According to \eqref{T_fb_G} and \eqref{distortion_all}, the rate-distortion trade-off under the Gaussian test channel model can be characterized by the following optimization problem
\begin{align}
    {(\rm P0):}~~\underset{\bm q}{\rm Minimize}~~&\Gamma_{\rm sum}({\bm q}) \notag\\
    {\rm Subject~ to}~~&\frac{B_k^{(G)}({\bm q}_k)}{\mu_k}\leq  \bar{T}_{\rm fb},~~\forall k \label{fb_con}.
\end{align}
where $\bar{T}_{\rm fb}$ is the given time constraint for quantization bit transmission. In other words, we aim to characterize given the feedback time constraint, what is the minimum quantization MSE.

\subsection{Quantization Bit Allocation}

In this section, we aim to design the quantization bit allocation solution of each user by solving problem $\rm (P0)$. According to \eqref{distortion_all}, which is due to the fact that users independently quantize their received signals, all the users can obtain their quantization bit allocation solutions in parallel. Specifically, the quantization bit allocation policy of user $k$ can be obtained by solving the following sub-problem:
\begin{align}
    ({\rm P1}\mbox{-}k):~~\underset{\{q_{k,i}\}}{\rm Minimize}~~&\Gamma_{k}({\bm q}_k) \notag\\
    {\rm Subject~ to}~~&B_k^{(G)}({\bm q}_k)\leq \mu_k \bar{T}_{\rm fb}.
\end{align}

In the following, we propose an efficient algorithm to solve problem $({\rm P1}\mbox{-}k)$ for user $k$, $k=1,\cdots,K$. Specifically, it can be shown that $\Gamma_k({\bm q}_k)$ given in \eqref{distortion} and $B_k^{(G)}({\bm q}_k)$ given in \eqref{B_k_G} are concave and convex functions over $q_{k,i}$'s, respectively. Therefore, the challenge is that we are minimizing a concave function, rather than a convex function. In this case, the majorization-minimization (MM) algorithm can be used to obtain a locally optimal solution \cite{sun2016majorization}.

MM is an iterative algorithm. Under the $s$-th iteration of the MM algorithm, we need to find a surrogate function of the objective function of problem $({\rm P1}\mbox{-}k)$. Because $\Gamma_k({\bm q}_k)$ is a concave function, its first-order Taylor expansion serves as its upper bound and can thus be used as its surrogate function. Specifically, given any point ${\bm q}_k^{(s)}=[q_{k,1}^{(s)},\cdots,q_{k,T}^{(s)}]^T$, the surrogate function of $\Gamma_k({\bm q}_k)$ can be set as
\begin{align}
    f\left({\bm q}_k\big|{\bm q}_k^{(s)}\right)&=\Gamma_k\left({\bm q}_k^{(s)}\right)+\left(\nabla \Gamma_k\left({\bm q}_k^{(s)}\right)\right)^T\left({\bm q}_k-{\bm q}_k^{(s)}   \right) \notag\\
    &\geq \Gamma_k\left({\bm q}_k\right),
\end{align}
where $\nabla\Gamma_k\left({\bm q}_k^{(s)}\right)$ is the derivative of $\Gamma_k({\bm q}_k)$ at ${\bm q}_k={\bm q}_k^{(s)}$ and given by
\begin{align}
    \nabla \Gamma_k\left({\bm q}_k^{(s)}\right)=\left[\frac{u_{k,1}^2}{\left(u_{k,1}+q_{k,1}^{(s)}\right)^2},\cdots,\frac{u_{k,T}^2}{\left(u_{k,T}+q_{k,T}^{(s)}\right)^2}\right]^T.
\end{align}

Then, under the $s$-th iteration of the MM algorithm, we need to solve the following problem
\begin{align}
    ({\rm P1}\mbox{-}k\mbox{-}s):~~\underset{\{q_{k,i}\}}{\rm Minimize}~~&f\left({\bm q}_k\big|{\bm q}_k^{(s)}\right) \notag\\
    {\rm Subject~ to}~~&B_k^{(G)}\left({\bm q}_k \right)\leq \mu_k \bar{T}_{\rm fb}.
\end{align}
Note that problem $({\rm P1}\mbox{-}k\mbox{-}s)$ is a convex optimization problem, and we can thus solve it globally based on the Lagrangian duality method. Specifically, the Lagrangian of problem $({\rm P1}\mbox{-}k\mbox{-}s)$ is expressed as
\begin{align}\label{Lagrange}
    L({\bm q}_k,\eta)=f\left({\bm q}_k\big|{\bm q}_k^{(s)}\right)+\eta\left(B_k^{(G)}({\bm q}_k)-\mu_k \bar{T}_{\rm fb}\right),
\end{align}
where $\eta\geq 0$ is the Language multiplier associated with the constraint of problem $({\rm P1}\mbox{-}k\mbox{-}s)$. The derivative of $L({\bm q}_k,\eta)$ over $q_{k,i}$ is expressed as
\begin{align}\label{deriv}
    \frac{\partial L({\bm q}_k,\eta)}{\partial q_{k,i}}=-\frac{\eta u_{k,i}}{(u_{k,i}+q_{k,i})q_{k,i}\ln2}+\frac{u_{k,i}^2}{\left(u_{k,i}+q_{k,i}^{(s)}\right)^2},~~\forall i.
\end{align}
By setting the derivative given in \eqref{deriv} as zero, it can be shown that given any $\eta\geq 0$, the Lagrangian given in \eqref{Lagrange} is minimized when
\begin{align}\label{q_ki_opt}
        \bar{q}_{k,i}(\eta)=\frac{1}{2}\left[ -u_{k,i}+\sqrt{u_{k,i}^2+\frac{4\eta\left(q_{k,i}^{(s)}+u_{k,i}\right)^2}{u_{k,i}\ln2}}\right], ~~\forall k,i.
\end{align}
Moreover, it can be shown that under the optimal solution to problem $({\rm P1}\mbox{-}k\mbox{-}s)$, the constraint should be satisfied with equality. Let $\eta^{\ast}$ denote the optimal Lagrange multiplier to problem $({\rm P1}\mbox{-}k\mbox{-}s)$. Then it follows that $\sum_{i=1}^T\log_2\left(1+u_{k,i}/\bar{q}_{k,i}(\eta^{\ast})\right)=\mu_k \bar{T}_{\rm fb}$. The solution $\eta^{\ast}$ to the above equation can be effectively obtained via the bisection method. Last, $\bar{q}_{k,i}(\eta^{\ast})$'s will be the optimal solution to problem $({\rm P1}\mbox{-}k\mbox{-}s)$.

After problem $({\rm P1}\mbox{-}k\mbox{-}s)$ is solved at the $s$-th iteration of the MM algorithm, we set ${\bm q}_k^{(s)}=[\bar{q}_{k,1}(\eta^{\ast}),\cdots,\bar{q}_{k,T}(\eta^{\ast})]^T$ and solve problem $({\rm P1}\mbox{-}k\mbox{-}s+1)$ for the $(s+1)$-th iteration of the MM algorithm. As shown in \cite{sun2016majorization}, under the MM algorithm, the objective value of problem $({\rm P1}\mbox{-}k)$ will decrease after each iteration, i.e., $\Gamma_k({\bm q}_k^{(s+1)})<\Gamma_k({\bm q}^{(s)})$, $\forall s$. Then, the MM algorithm will converge to a locally optimal solution to problem $({\rm P1}\mbox{-}k)$ \cite{sun2016majorization}.

Let ${\bm q}_k^{\ast}=[q_{k,1}^{\ast},\cdots,q_{k,T}^{\ast}]^T$ denote the solution to problem $({\rm P1}\mbox{-}k)$ obtained via the above MM algorithm. Then, we show how to determine the size of codebook $\mathcal{C}_{k,i}$ in \eqref{codebook}, i.e., $L_{k,i}$, $k=1,\cdots,K,i=1,\cdots,T$. According to \eqref{bits_ki}, the theoretical number of bits to quantize $y_{k,i}$ is
\begin{align}\label{Bki_G_opt}
    B_{k,i}^{(G)\ast}=\log_2\left(1+\frac{u_{k,i}}{{q}_{k,i}^{\ast}}\right),~~\forall k,i.
\end{align}
However, the above values may not be integer values. To satisfy the constraint in each problem $({\rm P1}\mbox{-}k)$, in practice, we can set the number of quantization bits of user $k$ at time sample $i$ as
\begin{align}
    B_{k,i}=\left\lfloor B_{k,i}^{(G)\ast}\right\rfloor, ~~\forall k,i,
\end{align}
where $\lfloor\cdot\rfloor$ denotes the floor function.
Thus, the size of $\mathcal{C}_{k,i}$ in \eqref{codebook} is given as
\begin{align}
    L_{k,i}=2^{B_{k,i}},~~\forall k,i.
\end{align}

\subsection{Tightness of the Gaussian Test Channel Approximation}

It was analytically shown in \cite{xia2006design} that the rate-distortion trade-off obtained from the Gaussian test channel model is a good approximation to that obtained from the Lloyd algorithm. In this sub-section, we provide a numerical example to verify this in our considered system.

In the numerical example, we assume that $D=16$, $M=12$ and $K=11$. The pilot transmission overhead $T$ is set as $32$ time samples and feedback transmission overhead $\bar{T}_{\rm fb}$ is set as $64$. Because quantization bit allocation can be performed independently over different users as shown in the previous sub-section, here we just focus on the rate-distortion trade-off of one user. In particular, we randomly generate $100$ bit allocation solutions, each satisfying constraint \eqref{fb_con}. Given each quantization bit allocation solution, we first calculate the MSE obtained under the Gaussian test channel model according to \eqref{distortion}, and then numerically calculate the channel estimation MSE obtained under the Lloyd algorithm based on Monte Carlo simulation.

\begin{figure}[t]
    \centering
    \includegraphics[scale=0.6]{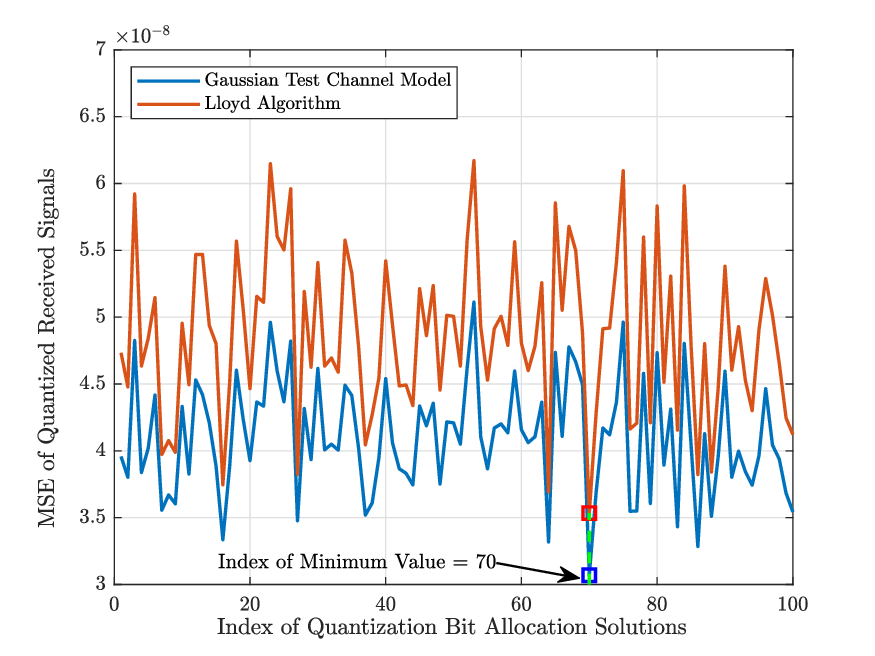}
    \caption{MSE comparison between the Gaussian test channel model and Lloyd algorithm under quantization bit allocation solutions.}
    \label{fig:test_vs_lloyd}
\end{figure}

The comparison between the above two MSEs under 100 quantization bit allocation solutions is given in Fig. \ref{fig:test_vs_lloyd}. It is observed that the gap between the MSEs achieved by the Gaussian test channel model and the Lloyd algorithm is very small. More importantly, for any two quantization bit allocation solutions, if one solution leads to a smaller MSE compared to another solution under the Gaussian test channel model, then usually this solution also leads to smaller MSE under the Lloyd algorithm. Moreover, among the $100$ quantization bit allocation solutions, the $70$th quantization bit allocation solution achieves the minimum MSE under both the Gaussian test channel model and the Lloyd algorithm. To summarize, the rate-distortion performance achieved by the Gaussian test channel model is a good approximation to that achieved by the Lloyd algorithm, and it is thus feasible to obtain the quantization bit allocation policies of different users by solving problems $({\rm P1}\mbox{-}k)$, $\forall k$. Note that similar observations have also been made in \cite{xia2006design}.

\vspace{-5pt}
\begin{figure}[t]
    \centering
    \includegraphics[scale=0.6]{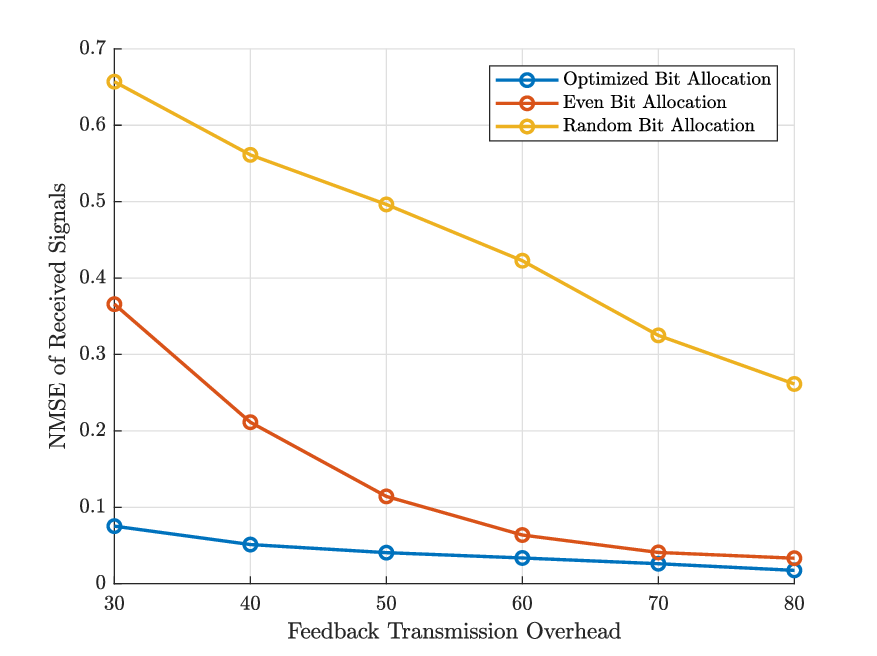}
    \caption{NMSE performance of received signals under different bit allocation solutions.}
    \label{fig:vs_bit_allos}
\end{figure}
\vspace{-5pt}
\subsection{Performance Comparison Under Different Bit Allocation Solutions}

In this sub-section, we show the gain of quantization bit allocation under our proposed algorithm. Two heuristic schemes are considered as the benchmark schemes. The first one is the even bit allocation scheme. Under this scheme, given the feedback overhead constraint $\bar{T}_{\rm fb}$, we set $B_{k,i}=\mu_k\bar{T}_{\rm fb}/T$, $\forall k,i$. The second one is the random bit allocation scheme. Under this scheme, for each user $k$, $B_{k,i}$'s are randomly generated and then normalized to satisfy the feedback overhead constraint $\bar{T}_{\rm fb}$.

Fig. \ref{fig:vs_bit_allos} shows the quantized received signals' normalized MSE (NMSE) over feedback transmission overhead achieved by different quantization bit allocation schemes. The NMSE for quantizing the received signals is defined as
\begin{align}
    {\rm NMSE}_y=\frac{\sum_{k=1}^K\mathbb{E}\left[\|\tilde{\bm y}_{k}-{\bm y}_{k}\|_2^2\right]}{\sum_{k=1}^K\mathbb{E}\left[\|{\bm y}_{k}\|_2^2\right]},
\end{align}
where $\tilde{\bm y}_k=[\tilde{y}_{k,1},\cdots,\tilde{y}_{k,T}]^T$, $\forall k$. The numbers of antennas at the BS, IRS sub-surfaces, and users are set as $M=16$, $D=20$, and $K=12$, respectively. The pilot transmission overhead $T$ is set as $40$ time samples and feedback transmission overhead $\bar{T}_{\rm fb}$ ranges from $30$ to $80$ time samples. It is observed that when the feedback overhead constraint is tight, our proposed quantization bit allocation scheme can achieve the best NMSE performance.

\section{Numerical Results}\label{sec:numerical}

In this section, we provide numerical results to demonstrate the advantages of our proposed ``quantize-then-estimate'' scheme for IRS-assisted communication. The BS-IRS channel ${\bm R}$ is modeled as a Rician fading channel with both the line-of-sight (LoS) deterministic component and the NLoS fading component:
\begin{align}
    {\bm R}=\sqrt{\frac{\kappa}{1+\kappa}}{\bm R}^{\rm LoS}+\sqrt{\frac{1}{1+\kappa}}({\bm C}^{\rm B})^{\frac{1}{2}}{\bm R}^{\rm NLoS}({\bm C}^{\rm I})^{\frac{1}{2}},
\end{align}
where $\kappa$ denotes the Rician factor set as $10$dB. ${\bm R}^{\rm LoS}$ denotes the LoS component in ${\bm R}$, ${\bm C}^{\rm B}\in\mathbb{C}^{M\times M}\succ{\bm 0}$ and ${\bm C}^{\rm I}\in\mathbb{C}^{D\times D}\succ{\bm 0}$ denote the BS transmit
correlation matrix and the IRS receive correlation matrix, respectively, and ${\bm R}^{\rm NLoS}\sim\mathcal{CN}({\bm 0},D\ell^{\rm BI}{\bm I})$ denotes the
i.i.d. Rayleigh fading component with
$\ell^{\rm BI}$ being the pass loss of ${\bm R}$. We assumed that ${\bm C}^{\rm B}$ and ${\bm C}^{\rm I}$ are generated based on the exponential correlation matrix model \cite{wang2020channel}. Next, the channel between the IRS and user $k$ is modeled as ${\bm t}_k=({\bm C}_k^{\rm I})^{\frac{1}{2}}\tilde{\bm t}_k$,
where ${\bm C}_k^{\rm I}\in\mathbb{C}^{D\times D}\succ{\bm 0}$ denotes the IRS transmit correlation matrix for user $k$, which also follows the exponential correlation matrix model, $\forall k$, and $\tilde{\bm t}_k\sim\mathcal{CN}({\bm 0},\ell^{\rm IU}_k{\bm I})$ follows the i.i.d. Rayleigh fading channel model with $\ell_k^{\rm IU}$ denoting the pass loss. Moreover, the pass loss of the BS-IRS channels ${\bm r}_d$'s and that of the IRS-user channels $t_{k,d}$'s are modeled as $\ell^{\rm BI}=\ell_0(\chi^{\rm BI}/\chi_0)^{\xi_1}$ and $\ell^{\rm IU}_k=\ell_0(\chi^{\rm IU}_k/\chi_0)^{\xi_2}$, $\forall k$, respectively, where $\ell_0$ meter (m) denotes the reference distance, $\ell_0=-20{\rm dB}$ denotes the path loss at the reference distance, $\chi^{\rm BI}$ and $\chi^{\rm IU}_k$ denote the distance between the BS and the IRS, and that between the IRS and user $k$, respectively, and $\xi_1=2.2$ and $\xi_2=2.1$ denote the path loss factors for ${\bm r}_d$'s and $t_{k,d}$'s, respectively. The distance between the BS and the IRS is set to be $100$ meters (m), and all the users are located in a circular region, whose center is $10$ m away from the IRS and $105$ m away from the BS and radius is $5$ m. The power spectrum density of the noise at the users is assumed to be $-169$ dBm/Hz, and the channel bandwidth is $1$ MHz. For the feedback transmission, it is assumed that each user employs 16QAM to modulate quantization bits, i.e., $\mu_k=4$, $\forall k$. Last, we use the NMSE as the metric to evaluate the channel estimation performance. Specifically, we define ${\bm H}_k=[{\bm g}_{k,1}^T,\cdots,{\bm g}_{k,D}^T]^T$ as the collection of user $k$'s cascaded channels, and $\hat{\bm H}_k=[\hat{\bm g}_{k,1}^T,\cdots,\hat{\bm g}_{k,D}^T]^T$ as the collection of their estimations, $k=1,\cdots,K$. Then, the overall NMSE for estimating all the cascaded channels is defined as
\begin{align}
    {\rm NMSE}=\frac{\sum_{k=1}^K\mathbb{E}\left[\|\hat{\bm H}_{k}-{\bm H}_{k}\|_2^2\right]}{\sum_{k=1}^K\mathbb{E}\left[\|{\bm H}_{k}\|_2^2\right]}.
\end{align}
In each numerical example, we conduct Monte Carlo simulation via $5000$ channel realizations to numerically obtain the NMSE performance.

In the following, we provide two benchmark schemes for channel estimation and feedback under our considered IRS-assisted systems and compare the performance of our proposed scheme over that of the benchmark schemes.

\begin{itemize}
    \item{\textbf{Benchmark Scheme 1}}: The first benchmark scheme is the conventional ``estimate-then-quantize'' scheme introduced in Section \ref{subsec:conv}. Under this scheme, each user $k$ applies the LMMSE technique on its received pilot signals given in \eqref{rev_com}  to estimate its own cascaded channels, denoted by $\bar{\bm g}_{k,d}$'s, and feeds back the estimated channels to the BS with codebook designed via the Lloyd algorithm.
    \item{\textbf{Benchmark Scheme 2}}: The second benchmark scheme is an improved ``estimate-then-quantize'' scheme. Under this scheme, each user $k$ still applies LMMSE technique to estimate its cascaded channels, denoted by $\bar{\bm g}_{k,d}=[\bar{g}_{k,d,1},\cdots,\bar{g}_{k,d,M}]^T$'s. However, for each IRS sub-surface $d$, only the reference user $k_d$ feeds back $\bar{\bm g}_{k_d,d}$ to the BS, and each user $k$ (including $k_d$) quantizes the sum of the estimated cascaded channel, i.e., $\sum_{m=1}^M \bar{g}_{k,d,m}$, and transmits the quantization bits to the BS. Then, the BS can estimate
    $\lambda_{k,d}$ for $k\neq k_d$ as
    \begin{align}
        \bar{\lambda}_{k,d}=\frac{\sum_{m=1}^M \bar{g}_{k,d,m}}{\sum_{m=1}^M \bar{g}_{k_d,d,m}}, ~~\forall d.
    \end{align}
    At last, the cascaded channels can be estimated based on \eqref{corr_group}. Note that compared to Benchmark Scheme 1, the main difference is that if a user is not a reference user, it merely feeds back the sum of its estimated channels thanks to \eqref{corr_group}. This can greatly reduce the feedback overhead. To characterize the overall overhead of Benchmark Scheme 2, let $\mathcal{S}_k=\{d:\forall d$ such that $k_d=k\}$ denote the set of sub-surfaces whose reference user is user $k$ based on criterion \eqref{k_d}, and $s_k=|\mathcal{S}_k|$ denote the number of times that user $k$ is selected as the reference user, $\forall k$. Therefore, each user $k$ needs to transmit $Ms_k$ samples about $\bar{\bm g}_{k,d}$, $\forall d\in\mathcal{S}_k$, and $D$ samples about $\bar{\lambda}_{k,d}$, $\forall d$. Then, the number of time samples for feedback transmission can be expressed as
    \begin{align}
        T_{\rm{fb,ben2}}=\underset{1\leq k \leq K}{\max}\left\{ \frac{(Ms_k+D)\lceil\log_2 L_{k}\rceil}{\mu_k} \right\},
    \end{align}
    and the minimum number of overall time samples for pilot and feedback transmission is thus
    \begin{align}\label{tol_ben2}
        T_{\rm{tol,ben2}}=MD+T_{\rm{fb,ben2}}.
    \end{align}
\end{itemize}

\begin{figure}[t]
    \centering
    \includegraphics[scale=0.6]{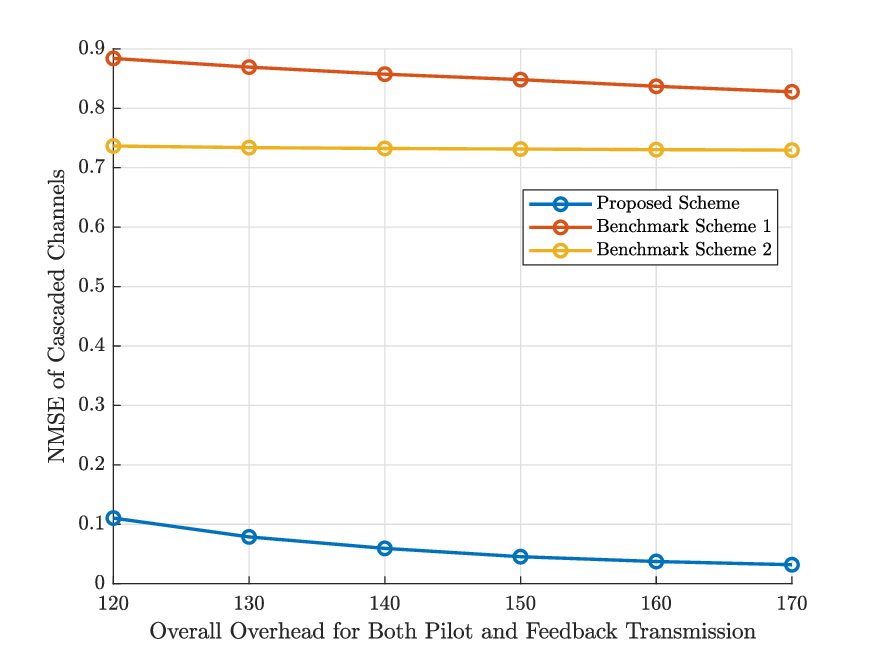}
    \caption{NMSE performance versus overall overhead of pilot and feedback transmission when overhead of pilot transmission is fixed as $40$ time samples.}
    \label{fig:fix_pilot}
    \vspace{-5pt}
\end{figure}

\begin{figure}[t]
    \centering
    \includegraphics[scale=0.6]{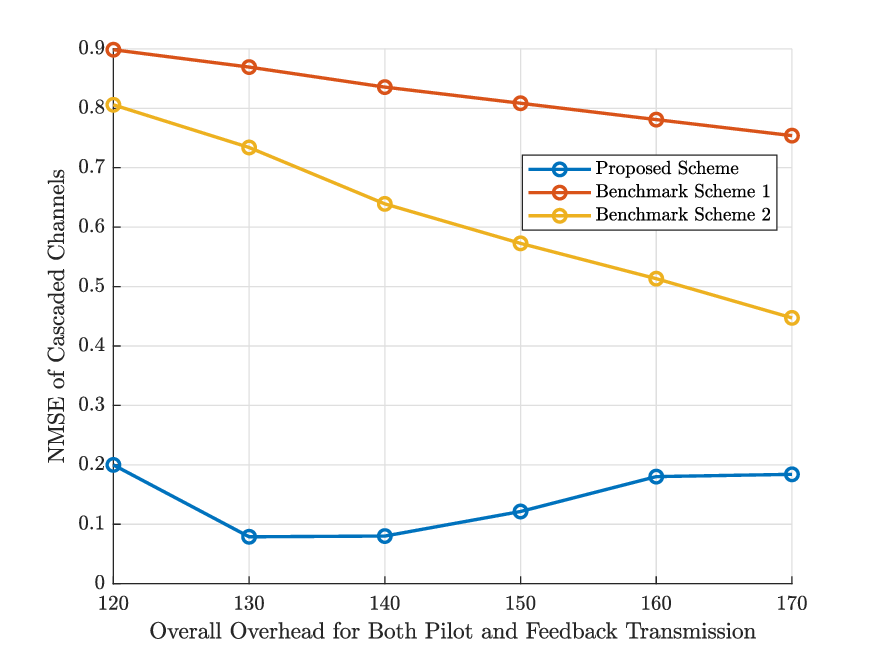}
    \caption{NMSE performance versus overall overhead of pilot and feedback transmission when overhead of feedback transmission is fixed as $90$ time samples.}
    \label{fig:fix_fb}
    \vspace{-5pt}
\end{figure}

In the numerical examples, the numbers of antennas at the BS, IRS sub-surfaces, and users are set as $M=12$, $D=16$, and $K=11$, respectively. The BS transmit power is $33$ dBm for all the time slots. Fig. \ref{fig:fix_pilot} shows the NMSE performance comparison between our proposed ``quantize-then-estimate'' scheme and the two benchmark schemes under the ``estimate-then-quantize'' approach. In this numerical example, the pilot transmission overhead is fixed as $40$ time samples, while the feedback transmission overhead ranges from $80$ to $130$ time samples such that the overall overhead for pilot and feedback transmission ranges from $120$ to $170$ time samples. It is observed that our proposed scheme shows a significant NMSE performance gain compared to the two benchmark schemes. For example, when the overall overhead is $170$ samples, the NMSE achieved by our proposed scheme is $0.0319$, which is much better than that achieved by the two benchmark schemes. This is because \eqref{corr_group} is leveraged to reduce both pilot and feedback transmission overhead. Note that Benchmark Scheme 2 shows a better performance than Benchmark Scheme 1 thanks to the reduction of feedback transmission overhead by exploiting \eqref{corr_group}. However, the performance of Benchmark Scheme 2 is much worse than that of our proposed scheme. This is because Benchmark Scheme 2 cannot utilize \eqref{corr_group} to reduce the pilot transmission overhead. Specifically, given $M=12$, $D=16$, and $K=11$, the minimum numbers of time samples required by Benchmark Scheme 2 and our proposed scheme are $192$ and $32$, respectively. Therefore, when the pilot transmission overhead is fixed as $40$ time samples, the users cannot accurately estimate their cascaded channels under Benchmark Scheme 2.

Next, we set the feedback transmission overhead as $90$ time samples, and the pilot transmission overhead ranges from $30$ to $80$ time samples, i.e., the overall overhead ranges from $120$ to $170$ time samples. Fig. \ref{fig:fix_fb} shows the performance comparison among different schemes. Under our proposed scheme, it is observed that as the overall overhead increases, the NMSE shows a “first-drop-then-rise”trend, while the optimal NMSE is achieved when the pilot transmission overhead is $40$ time samples, i.e., the overall overhead is $130$ time samples. Note that as the pilot transmission overhead increases, the BS can estimate the channels based on more received signals, but needs to quantize more pilot signals given the number of quantization bits. When the pilot transmission overhead is small, pilot transmission is the bottleneck to limit the channel estimation performance at the BS, and it is beneficial to increase the pilot transmission overhead. However, when the pilot transmission overhead is large enough, the BS has enough pilot signals to estimate the channels, and it is not a good idea to keep increasing the pilot transmission overhead because this will reduce the number of bits to quantize each pilot sample. This indicates that the pilot transmission overhead should be carefully designed. For the two benchmarks, increasing pilot transmission overhead always results in a decreasing NMSE because when the users feed back their channels, the amount of feedback is independent of the number of pilot signals, and the users can quantize better-estimated channels given the same number of quantization bits.

\begin{figure}[t]
    \centering
    \includegraphics[scale=0.6]{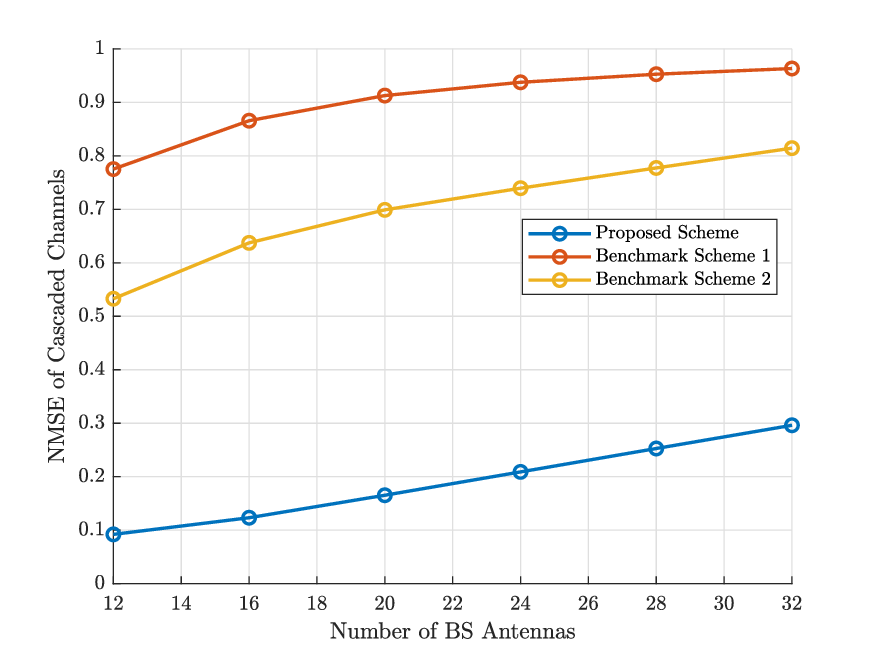}
    \caption{NMSE performance versus number of BS antennas.}
    \label{fig:vs_antenna}
    \vspace{-5pt}
\end{figure}

Moreover, Fig. \ref{fig:vs_antenna} shows the estimated channels' NMSE versus the number of BS antennas, with $D=16$ and $K=11$. The pilot transmission overhead is $64$ time samples and the feedback transmission overhead is $100$ time samples. It is observed that when the number of BS antennas increases, the NMSE increases much more slowly under our proposed scheme than the two benchmark schemes. This is attributed to exploiting \eqref{corr_group} to reduce both pilot and feedback transmission overhead. Specifically, every additional BS antenna increases $D/K\approx1.4545$ time samples of pilot and feedback transmission overhead, respectively, under our proposed scheme. While it causes $16$ time samples of pilot and feedback transmission overhead, respectively, under Benchmark Scheme 1; and $D=16$ time samples of pilot transmission overhead and $D/K\approx1.4545$ time samples of feedback transmission overhead under Benchmark Scheme 2. 

\begin{figure}[t]
    \centering
    \includegraphics[scale=0.6]{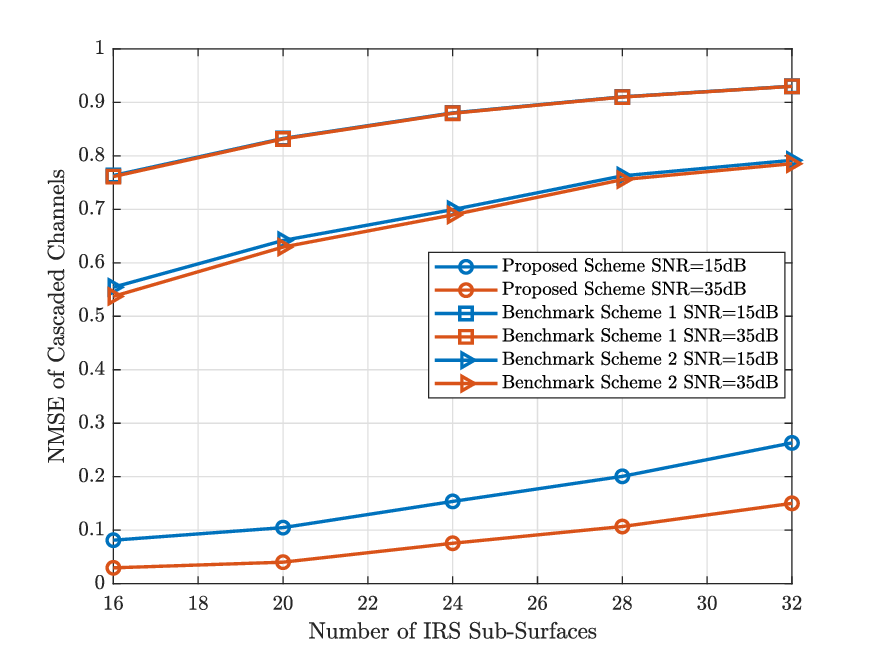}
    \caption{NMSE performance versus SNR and number of IRS sub-surfaces.}
    \label{fig:vs_IRS}
\end{figure}

Last, Fig. \ref{fig:vs_IRS} shows the estimated channels' NMSE of our proposed scheme versus different numbers of IRS sub-surfaces and signal-to-noise ratios (SNRs), where $M=16$, $K=11$, the pilot transmission overhead is set as $80$ time samples, and the feedback transmission overhead is set as $160$ time samples. It is observed that over different SNRs and numbers of IRS sub-surfaces, our proposed scheme performs better than the benchmark schemes.

\section{Conclusion}\label{sec:conclusion}

In this paper, we studied downlink CSI acquisition in FDD IRS-assisted communication systems. Motivated by the correlated channels among different users, we proposed a novel ``quantize-then-estimate" protocol for reducing the overhead in both pilot transmission and feedback transmission. Specifically, all the users first quantize their received pilot signals and then transmit the quantization bits to the BS. After de-quantizing all the user's received signals, the BS can thus leverage the correlation embedded in users' cascaded channels to perform channel estimation. We designed efficient methods for each user to allocate the quantization bits over time and quantize the signals based on the carefully devised codebook, and for the BS to perform the LMMSE technique for estimating the channels based on the quantized signals. Moreover, we analytically characterized the minimum overhead for pilot transmission and feedback transmission under our proposed ``quantize-then-estimate" protocol and demonstrated the significant overhead reduction compared to the conventional ``estimate-then-quantize" protocol. Our results open up a new solution for low-overhead communication in IRS-assisted systems.

\appendix \label{apx}

We prove the theorem in two cases: 1) $K\geq D$ and 2) $K<D$. In the case of $K\geq D$, we first prove that there exists a unique solution to \eqref{tilde_y2} only if $\tau_2\geq M-1$. Define
\begin{equation}\label{eta}
  \eta_{d,i}=\sqrt{p}{\bm g}_{k_d,d}^{T}{\bm x}_i, ~~d=1,\cdots, D, i=\tau_1+1,\cdots,\tau_1+\tau_2.
\end{equation}
Then the received pilot signals of user $k$ at time sample $i$ can be expressed as
\begin{equation}
    \tilde{y}_{k,i}=\sum_{d=1}^{D}\phi_{d,i}\lambda_{k,d}\eta_{d,i}, ~~\forall k, i=\tau_1+1,\cdots,\tau_1+\tau_2,
\end{equation}
Define ${\bm\lambda}=[{\bm \lambda}_1,\cdots,{\bm \lambda}_{D}]$, then the overall received pilots at time sample $i$ can be re-written as
\begin{align}\label{y2_eta}
    \tilde{\bm y}^{(2)}_i=[\tilde{y}_{1,i},\cdots,\tilde{y}_{K,i}]^T={\bm\lambda}{\bm \Gamma}_i,~~i=\tau_1+1,\cdots,\tau_1+\tau_2
\end{align}
where ${\bm \Gamma}_i=[\phi_{1,i}\eta_{1,i},\cdots,\phi_{D,i}\eta_{D,i}]^T$. We set $\phi_{d,i}=1, \forall d,i$, then ${\bm \Gamma}_i=[\eta_{1,i},\cdots,\eta_{D,i}]^T$.
With $\lambda_{k,d}$'s estimated in Step 1, there exist $D$ variables $\eta_{d,i}$'s and $K$ linear equations as given in \eqref{y2_eta}. As a result, in the case of $K\geq D$, $\eta_{d,i}$'s can be perfectly estimated. Then, with the knowledge of $\eta_{d,i}$'s and ${\alpha}_{k_d,d}$'s, we can estimate ${\bm g}_{k_d,d}$'s based on \eqref{alpha} and \eqref{eta} from the following equations
\begin{align}\label{eta&alpha}
  &\left[{\alpha}_{k_d,d},\eta_{d,\tau_1+1},\cdots,\eta_{d,\tau_1+\tau_2}\right]^T \notag\\
  &=\sqrt{p}\left[{\bm x},{\bm x}_{\tau_1+1},\cdots,{\bm x}_{\tau_1+\tau_2}
  \right]^T{\bm g}_{k_d,d}, \notag\\
  &d=1,\cdots, D,
\end{align}
which characterizes a linear system with $MD$ variables and $(\tau_2+1)D$ equations. Therefore, a unique solution to \eqref{eta&alpha} exists only when the number of equations is no smaller than the number of variables, i.e. $\tau_2\geq M-1$.

Next, we show that if $\tau_2=M-1$, there always exists a unique solution to \eqref{tilde_y2} in the case of $K\geq D$. Specifically, since ${\bm\lambda}_d$'s are linearly independent with each other with probability one, $\eta_{d,i}$'s can be perfectly estimated based on \eqref{y2_eta} as
\begin{align}\label{es_eta}
  {\bm\Gamma}_i={\bm\lambda}^{\dagger}\tilde{\bm y}^{(2)}_i,~~i=\tau_1+1,\cdots,\tau_1+M-1.
\end{align}
Then we set $\bm x=[1,\cdots,1]^T$, and ${\bm x}_{\tau_1+1},\cdots,{\bm x}_{\tau_1+M-1}$ as the $2$ to $M$ columns of a $M \times M$ DFT matrix. With the knowledge of $\eta_{d,i}$'s and $\alpha_{k_d,d}$'s, $d=1,\cdots,D,i=\tau_1+1,\cdots,\tau_1+M-1$, there exists a unique solution to \eqref{eta&alpha}, equivalently to \eqref{tilde_y2} given as follows
\begin{align}\label{es_g_K>D}
    &{\bm g}_{k_d,d}=\sqrt{p}[{\bm x},{\bm x}_{\tau_1+1},\cdots,{\bm x}_{\tau_1+M-1}]^* \notag\\
    &\times[{\alpha}_{k_d,d},\eta_{d,\tau_1+1},\cdots,\eta_{d,\tau_1+M-1}]^T, ~~d=1,\cdots,D.
\end{align}

In the case of $K<D$, since the number of variables and equations in \eqref{tilde_y2} are $MD$ and $\tau_2K+D$, respectively, there exists a unique solution to \eqref{tilde_y2} only if the number of equations is no smaller than that of variables, i.e., $\tau_2\geq\lceil\frac{(M-1)D}{K}\rceil$.

Next, we show that when $\tau_2=\lceil\frac{(M-1)D}{K}\rceil$, there always exists a solution to \eqref{tilde_y2} in the case of $K<D$. Specifically, we first set the pilot signal in Step 1 as an all-one vector, i.e., ${\bm x}=[1,\cdots,1]^T$, and the $M$-th pilot signal equal to zero in Step 2, i.e., ${\bm x}_{i,M}=0, i=\tau_1+1, \cdots, \tau_1+\tau_2$. Then the other pilot signals, i.e., ${\bm x}_{i,m}, m=1,\cdots,M-1$, as well as the IRS reflecting coefficients, i.e., $\phi_{d,i}, d=1,\cdots,D, i=\tau_1+1,\cdots,\tau_1+\tau_2$, are set in the same way as Theorem $2$ in \cite{wang2020channel}. Then, we construct a new matrix $\bar{\bm \Theta}\in\mathbb{C}^{(\tau_2K+D)\times MD}$ by putting the $[(d-1)M+m]$-th column of $\bm \Theta$ into the $[(m-1)D+d]$-th column of $\bar{\bm \Theta}$, $\forall m,d$. Since changing the order of columns of a matrix does not change its rank, i.e., ${\rm rank}(\bar{\bm \Theta})={\rm rank}(\bm \Theta)$, in the following, we show that under the above construction, we have ${\rm rank}(\bar{\bm \Theta})=MD$ when $\tau_2=\lceil\frac{(M-1)D}{K}\rceil$. Specifically, $\bar{\bm \Theta}$ can be re-expressed as follows:
\begin{equation}\label{Theta_bar}
  \bar{\bm \Theta}=\left[
  \begin{array}{cc}
    \bar{\bm\Theta}_{s} & {\bm O}_{\tau_2K \times D}\\
    \{{\bm I}_D\}_{M-1} & {\bm I}_D
  \end{array}\right],
\end{equation}
where $\bar{\bm\Theta}_{s}$ is the first $\tau_2K$ rows and first $(M-1)D$ columns of $\bar{\bm\Theta}$, ${\bm O}_{\tau_2K\times D}$ is an all-zero matrix with dimension $\tau_2K\times D$, ${\bm I}_D$ is the identity matrix of dimension $D$, and $\{{\bm I}_D\}_{M-1}=[{\bm I}_D,\cdots,{\bm I}_D]\in\mathbb{C}^{D\times (M-1)D}$. According to Theorem $2$ in \cite{wang2020channel}, ${\rm rank}(\bar{\bm \Theta}_s)=(M-1)D$ when $\tau_2=\lceil\frac{(M-1)D}{K}\rceil$. Next, we derive the rank of $\bar{\bm\Theta}$. It is observed from \eqref{Theta_bar} that each of the first $\tau_2 K$ rows of $\bar{\bm \Theta}$, whose last $D$ elements are all zero, is linearly independent of the last $D$ rows of $\bar{\bm\Theta}$, i.e., $\{{\bm I}_D\}_{M}$. In other words, the row space defined by the first $\tau_2 K$ rows in $\bar{\bm\Theta}$ does not intersect with that defined by the last $D$ rows in $\bar{\bm\Theta}$. In this case, ${\rm rank}(\bar{\bm \Theta})={\rm rank}([\bar{\bm \Theta}_s~~{\bm O}_{\tau_2K\times D}])+{\rm rank}(\{{\bm I}_D\}_{M})=MD$ \cite{Rank}. Therefore, for the case of $K<D$, when $\tau_2=\lceil\frac{(M-1)D}{K}\rceil$, there exists a unique solution to \eqref{tilde_y2} given by
\begin{equation}\label{es_g_K<D}
  {\bm g}={\bm\Theta}^{\dagger}\tilde{\bm y}^{(2)}.
\end{equation}

Theorem $1$ is thus proved.

\bibliographystyle{IEEEtran}
\bibliography{reference}

\end{document}